\documentclass[aps,prb,floatfix,superscriptaddress,amsmath,amssymb,twocolumn]{revtex4-1}
\usepackage[latin1]{inputenc} \usepackage{amsmath} \usepackage{amssymb}
\usepackage{graphicx} \usepackage{graphics} \usepackage{color}
\usepackage{multirow} \usepackage{subfigure}

 \def \ba {\begin{eqnarray}} \def \ea {\end{eqnarray}}
\newcommand{\nn}{\nonumber}

\newcommand{\figu}[2][0.3]{
\raisebox{-2.5ex}{\includegraphics[scale=#1]{#2}} }

\newcommand{\fifi}[3][0.3]{
\raisebox{#2ex}{\includegraphics[scale=#1]{#3}} }

\newcommand{\sumastar}{%
\raisebox{0.0ex}{\includegraphics[scale=0.08]{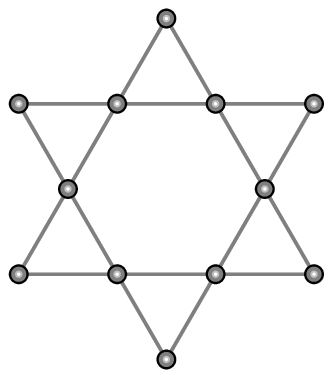}} }

\newcommand{\figket}[2][0.3]{\Bigl|\,
\raisebox{-0.9ex}{\includegraphics[scale=#1]{#2}}\, \Bigr\rangle}

\newcommand{\figbra}[2][0.3]{\Bigl\langle \, 
\raisebox{-0.9ex}{\includegraphics[scale=#1]{#2}}\,  \Bigr|}

\newcommand{\configA}{\Bigl|\,%
\raisebox{-1.0ex}{\includegraphics[scale=0.3]{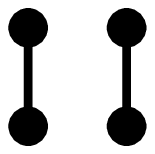}}\,%
\Bigr\rangle}

\newcommand{\configB}{\Bigl|\,%
\raisebox{-1.0ex}{\includegraphics[scale=0.3]{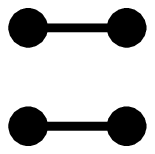}}\,%
\Bigr\rangle}

\newcommand{\configHuno}{\Bigl|\,%
\raisebox{-1.0ex}{\includegraphics[scale=0.3]{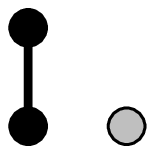}}\,%
\Bigr\rangle}

\newcommand{\configHtres}{\Bigl|\,%
\raisebox{-1.0ex}{\includegraphics[scale=0.3]{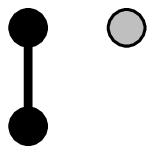}}\,%
\Bigr\rangle}

\newcommand{\configHcinco}{\Bigl|\,%
\raisebox{-1.0ex}{\includegraphics[scale=0.3]{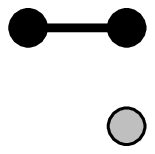}}\,%
\Bigr\rangle}

\newcommand{\configHsiete}{\Bigl|\,%
\raisebox{-1.0ex}{\includegraphics[scale=0.3]{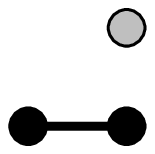}}\,%
\Bigr\rangle}

\newcommand{\braHdos}{\Bigl\langle \,%
\raisebox{-1.0ex}{\includegraphics[scale=0.3]{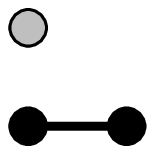}}\,%
\Bigr|}

\newcommand{\braHcuatro}{\Bigl\langle \,%
\raisebox{-1.0ex}{\includegraphics[scale=0.3]{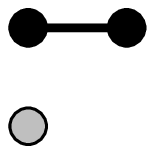}}\,%
\Bigr|}

\newcommand{\braHseis}{\Bigl\langle \,%
\raisebox{-1.0ex}{\includegraphics[scale=0.3]{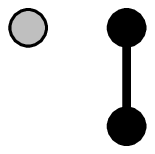}}\,%
\Bigr|}

\newcommand{\braHocho}{\Bigl\langle \,%
\raisebox{-1.0ex}{\includegraphics[scale=0.3]{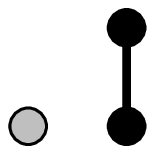}}\,%
\Bigr|}

\newcommand{\braA}{\Bigl\langle \,%
\raisebox{-1.0ex}{\includegraphics[scale=0.3]{estadoA.eps}}\,%
\Bigr|}

\newcommand{\braB}{\Bigl\langle \,%
\raisebox{-1.0ex}{\includegraphics[scale=0.3]{estadoB.eps}}\,%
\Bigr|}

\newcommand{\Vhex}{V_{
\raisebox{-0.5ex}{\includegraphics[scale=0.1]{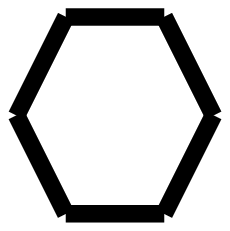}}}}

\newcommand{\hexagono}{
\raisebox{-0.5ex}{\includegraphics[scale=0.1]{hexagon.eps}}}

\begin{document}

\title{Statistics of holes and nature of superfluid phases in
Quantum dimer models.}

\author{C.A. Lamas} \affiliation{Laboratoire de Physique Th\'eorique,
IRSAMC, CNRS and Universit\'e de Toulouse, UPS, F-31062 Toulouse,
France}

\author{A. Ralko} \affiliation{Institut N\'eel, CNRS and Universit\'e
Joseph Fourier, F-38042 Grenoble, France}

\author{M. Oshikawa} \affiliation{Institute for Solid State Physics,
University of Tokyo, Kashiwa 277-8581, Japan}

\author{D. Poilblanc} \affiliation{Laboratoire de Physique Th\'eorique,
IRSAMC, CNRS and Universit\'e de Toulouse, UPS, F-31062 Toulouse,
France}

\author{P. Pujol} \affiliation{Laboratoire de Physique Th\'eorique,
IRSAMC, CNRS and Universit\'e de Toulouse, UPS, F-31062 Toulouse,
France}

\begin{abstract}
Quantum Dimer Models (QDM) arise as low energy effective models for
frustrated magnets.  Some of these models have proven successful in
generating a scenario for exotic 
spin liquid phases with deconfined spinons. Doping, i.e. the introduction of mobile holes, 
has been considered within the QDM framework and partially studied. A fundamental
issue is the possible existence of a superconducting phase in such systems and its properties.  
For this purpose,
the question of the statistics of the mobile holes (or ``holons") shall be addressed first.  
Such issues are studied in details in this paper 
for generic doped QDM defined on the most common two-dimensional lattices
(square, triangular, honeycomb, kagome,...) and involving general resonant loops.  
We prove a general ``statistical
transmutation'' symmetry of such doped QDM 
by using composite operators of dimers and holes.  This exact
transformation enables to define duality equivalence classes (or families) of
doped QDM, and provides the analytic framework to analyze dynamical
statistical transmutations.  We discuss various possible superconducting
phases of the system. In particular, the possibility of
an exotic superconducting phase originating from the condensation of 
(bosonic) charge-$e$ holons is examined.  A numerical evidence of such
a superconducting phase is presented in the case of the triangular lattice,
by introducing a novel gauge-invariant holon Green's function.
We also make the
connection with a Bose-Hubbard model on the kagome lattice which gives rise,
as an effective model in the limit of strong interactions,
to a doped QDM on the triangular lattice.
\end{abstract}

\maketitle

\section{Introduction}

In 1987 Anderson \cite{Anderson87} suggested the strange behavior of
cuprate materials between the superconducting dome and the magnetically
ordered insulating phase could be described by a Resonating Valence Bond
(RVB) state in which preexisting magnetic singlet pairs of the
insulating state become charged superconducting pairs when the insulator
is doped.  Just one year later appeared the first effective model in
which the magnetic degrees of freedom are disregarded in favor of the
more pertinent singlet degrees of freedom \cite{RK}.
This is nothing but the Quantum Dimer Model (QDM).
However, it was soon realized\cite{HAFsquareNeel} that the
groundstate of the undoped $S=1/2$ Heisenberg antiferromagnet
on a square lattice exhibits a long-range N\'{e}el order
in contrast to the initial expectation based on the RVB picture.
Furthermore, the QDM on a square lattice was also found
to have only
gapped crystalline phases but no evidence of an RVB spin liquid phase in a finite region of the
phase diagram.\cite{RVB-square}
It is still possible to argue that, even though the
undoped antiferromagnet has the N\'{e}el order,
the RVB picture gives a better theoretical starting point
once the system is doped with holes.
However, it would be natural to ask if the RVB spin liquid
phase can be realized in undoped magnetic system with only
short-range interaction.


One may expect that magnetic frustration would favor RVB state over the
N\'{e}el phase.
Thus, over the time, the main interest in
QDM was shifted
from the original motivation of the application to high $T_c$
superconductivity,
to the effects of frustration.
However, clear confirmation of a RVB phase remained elusive for
a rather long time.
A breakthrough in the study of QDM was due to Moessner and Sondhi
\cite{Moessner_prl_2001_Z2} who showed that a simple QDM defined in the
triangular lattice exhibit a disordered phase which, recalling that these dimer
models are supposed to be effective models for frustrated magnets, can
be considered as an explicit example of the RVB
spin liquid phase.
It was also recognized that, the RVB spin liquid phase
is a topologically ordered phase with a nontrivial topological
degeneracy of the groundstates.\cite{Bonesteel1989}
In fact, the RVB spin liquid phase is essentially identical
to the $\mathbb{Z}_2$ topological phase which was introduced in a
completely different context of quantum
information processing.\cite{Kitaev-toric}
The QDM is generally not exactly eqivalent to an antiferromagnetic
Hamiltonian defined in terms of quantum spins.
However, the projection from a magnetic system to a QDM was performed successfully in 
Heisenberg antiferromagnets on frustrated lattice, such as
the square lattice with strong
enough second and/or third neighbors couplings \cite{Square_j1_j2_j3,
Heisenberg->square} or the kagome lattice \cite{Heisenberg->kagome}.
These suggest that the QDM may well represent phases without magnetic order in antiferromagnets. In fact, very recently, frustrated Heisenberg antiferromagnets on kagome and other lattices are reported to be in the RVB spin liquid phase ($\mathbb{Z}_2$ topological phase) by several authors.\cite{Kagome_Z2_Science2011,Kagome_Z2_Jiang2012,Kagome_Z2_Depenbrock2012}

Now that the existence of the RVB spin liquid phase appears to be confirmed in QDMs as well as in antiferromagnets, 
the issue of superconductivity in doped spin liquids becomes a more pressing question. 
This isssue started in fact to be investigated shortly after the apearence of QDM \cite{Fradkin+Kivelson}
Doping of an RVB spin liquid is expected
to induce a novel type of elementary excitations called holon.
A holon, carrying electric charge $e$ but no spin, appears
as a result of fractionalization, namely deconfinement of
fractionalized excitations.
Indeed, topological degeneracy of the undoped RVB spin liquid
is known to be intimately connected to the fractionalization
phenomenon\cite{OS-fractionalization}.

Superconductivity may be realized if the holons condense.
At least naively, one may expect that the resulting superconductor
is an exotic one due to condensation of charge-$e$ holons,
instead of usual charge-$2e$ Cooper pairs.
A fundamental issue in this problem is the statistics of
the holon. For the holons to condense without forming pairs,
they must be bosons.
However, it should be noted that transmutation of
the statistics\cite{read, Kivelson_1989} is possible.
Namely,
the statistics of holons as elementary excitations appearing
in the low-energy limit can be different from
the statistics assigned to holons in
the microscopic model.

In this paper we address the issue of the statistics of holes and
its interplay with possible superconducting phases in doped QDMs.
In a recent work\cite{stat1} it was shown that
a QDM with fermionic (at microscopic level) holes
is equivalent to another QDM with bosonic holes.
Because of the equivalence, the statistics of the holon
as a physical elementary excitation must be the same
for either representation.
This proves the existence of a dynamical statistical transmutation
in the system. 
In this paper we study in more
details the statistical transmutation in QDM and give a simple and
efficient method to obtain the relation between the QDMs
with fermionic and bosonic representation of the holes.

In section~\ref{sec.gauge}
we introduce a second quantization notation for QDM
Hamiltonians and show the gauge symmetry associated with them. In
section~\ref{sec.composite} we present
the composite particle representation of QDM
Hamiltonians which is the key ingredient to show the exact equivalence
between a QDM with bosonic and another QDM with fermionic holes. This
equivalence is shown for a generic flipping term defined in any kind of
lattices. The result, which relies in an orientation prescription of the
bonds in the lattice considered, is totally generic and can then be
applied to any QDM defined in the most common lattices. It is important
to stress that the method used here differs considerably with the one
used in Ref.~\onlinecite{stat1} where a two-dimensional version of the
Jordan-Wigner (JW) transformation was used.
As we explain in the conclusion, it is important to point out the numerous 
advantages of this new representation
compared to the JW transformation. 
In section~\ref{sec.classification} we argue how the
modification of the orientation prescription can be interpreted as a
simple gauge transformation in the QDM Hamiltonian. We then apply the
general result of the statistical transmutation obtained in
section~\ref{sec.composite}
to generic QDM Hamiltonians defined on the square, triangular, hexagonal
and kagome lattices.
Section~\ref{sec.numerical} is devoted to numerical investigation
of four inequivalent QDM defined on the triangular lattice.
In particular, we identify an exotic superconductor phase
due to condensation of holons with charge $e$,
measuring the gauge-invariant Green's function of a single holon.
In section~\ref{sec.BH} we discuss an explicit realization
of one of the QDM discussed in section~\ref{sec.numerical}.
It is obtained as a low energy strong interaction limit of a
Bose-Hubbard model on the kagome lattice. The number of bosons is
directly related to the doping, or number of holes, in the resulting QDM
on the triangular lattice. Section~\ref{sec.conclusions}
is devoted to the discussion of
our results. We also include as an appendix the derivation of the
statistical transmutation for a generic QDM on the kagome lattice using
the Jordan-Wigner transformation. Of course the result is consistent
with the one obtained with the composite particle representation
obtained in section~\ref{sec.composite},
but allows a better understanding of the
connection between these two different methods.

\section{The Hamiltonian and its gauge symmetries. }
\label{sec.gauge}

We start with a doped quantum dimer model on a two-dimensional lattice.
 To fix the ideas, we work here with the Hamiltonian defined on the
 square lattice but all the arguments remain valid for any two
 dimensional lattice.  We write the Hamiltonian as
\ba H=H_{J}+H_{V}+H_{t} \label{eq:H1} \ea
with
\ba \nn H_{J}&=&-J\sum_{\Box} \left[\configA \braB + \hbox{H.C}\right]\\
\nn H_{V}&=&V\sum_{\Box} \left[\configA \braA + \configB \braB \right]\\
\nn H_{t}&=&-t\sum_{\Box}\left\{ \configHuno \braHdos + \configHtres
\braHcuatro \right.\\ \nn &+& \left. \configHcinco \braHseis +
\configHsiete \braHocho + \hbox{H.C} \right\}, \ea
where the sums are over all the smallest resonant plaquettes on the
lattice (for the square lattice these are the squares).
In a second quantized formalism we assume that dimer configurations are
 created by spatially symmetric dimer operators $b^{\dagger}_{i,j}$ and
 holes are created by bosonic operators $a^{\dagger}_{k}$. Then, we can
 re-write the Hamiltonian as:
\ba H_{J}&=&-J\sum_{\square}\left\{ b^{\dagger}_{i,j}b^{\dagger}_{k,l}
b_{j,k}b_{l,i}+\hbox{H.c.}  \right\}\\
H_{V}&=&V\sum_{\square}\left\{ b^{\dagger}_{i,j}b^{\dagger}_{k,l}
b_{i,j}b_{k,l}
+ b^{\dagger}_{j,k}b^{\dagger}_{l,i} b_{j,k}b_{l,i} \right\}\\
H_{t}&=&-t\sum_{i}\left\{ b^{\dagger}_{i,j}b_{j,k} a^{\dagger}_{k}a_{i} +
\hbox{H.c} \right\}.  \label{hoping_t} \ea
In the last equation, the indices correspond to the labeling of the sites of a square 
plaquette as in Figure \ref{fig:index}. 
In our previous conventions, dimer configurations are represented by
 spatially symmetric operators $b^{\dagger}_{i,j}$ satisfying:
\ba \label{eq:commut_b} \left[b_{i,j} , b^{\dagger}_{k,l}\right] &=&
\delta_{i,k}\delta_{j,l}+\delta_{i,l}\delta_{j,k}\\\nonumber
\left[ b_{i,j} , b_{k,l}\right] &= & \left[b^{\dagger}_{i,j} ,
b^{\dagger}_{k,l}\right] = 0 .  \ea
The boson operator $a^{\dagger}_{i}$ creates a hole in the site $i$ and
satisfies
\ba \label{eq:commut_a} \left[a_{i} , a^{\dagger}_{j}\right] &=&
\delta_{i,j}\\\nonumber
\left[ a_{i} , a_{j}\right] &= & \left[a^{\dagger}_{i} ,
a^{\dagger}_{j}\right] = 0.
\ea
The operators  $a$ and $b$ commutes one with each
other \ba \label{eq:commut_ab}
\left[ a_{i} , b_{j}\right] = \left[a^{\dagger}_{i} ,
b^{\dagger}_{j}\right] =\left[a^{\dagger}_{i} , b_{j}\right]= 0.  \ea
We introduce in the model a constraint on the number of dimers and holes
which warrants that at each site of the lattice there is either one and
only one hole or one and only one dimer arriving to it:
\ba \label{eq:constraint}
a^{\dagger}_{i}a_{i}+\sum_{z=\pm\hat{e}_{1},\pm\hat{e}_{2}}b^{\dagger}_{i,i+z}b_{i,i+z}=1.
\ea
Of course this constraint implies, among others, that the holes have to
be considered as hard core bosons. It is important to notice that the
Hamiltonian has the following $U(1)$ gauge symmetry:
\ba \label{eq:gauge1} a_{j} & \rightarrow &
e^{i\xi_{j}}a_{j}\\\label{re:gauge2} b_{j,k} & \rightarrow &
e^{i(\xi_{j}+\xi_{k})}b_{j,k} \ea
where $\xi_{i}$ is an angle. This invariance can be exploited to prove
the statistical transmutation symmetry in some two dimensional systems
by means a Jordan-Wigner transformation on the holon
operators\cite{stat1}.  In the following we present an alternative
description for the doped QDM using composite operators which
allows us to
understand in a different way the equivalence between a model with
bosonic holes and one with fermionic holes. For doing this, we have
first to make a choice of a given orientation prescription for the
bonds in the lattice.

\begin{figure}[t!]
 \begin{center}
\includegraphics*[angle=0,width=0.4\textwidth]{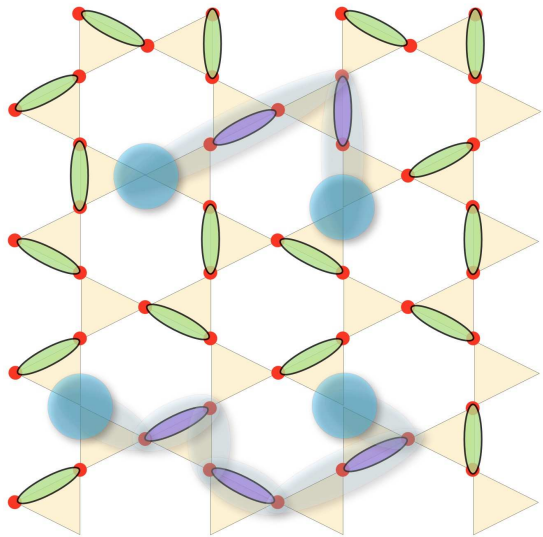}
\end{center}
\caption{ \label{fig:soup} (color online) Schematic snapshot of a doped
``dimer liquid". Each site is occupied by either a (single) dimer or a
hole (empty site). }
\end{figure}

\section{The ``composite'' representation for the QDM}
\label{sec.composite}

\subsection{Composite particles}
\label{Comp-part}

In order to prove the equivalence between a QDM Hamiltonian with bosonic
holes and another Hamiltonian with fermionic holes, we propose a
different formulation for the QDM. This formulation is done in terms of
composite particles by defining the operator \ba
B_{i,j}=b_{i,j}a^{\dagger}_{i}a^{\dagger}_{j}.  \ea
This operator destroys a dimer between sites $i$ and $j$ and creates two
holes at the same sites.  Let's call $\mathcal{H}_{c}$ the subspace of
states that satisfies the constraint (\ref{eq:constraint}).  For a given
state $|\psi\rangle \in \mathcal{H}_{c}$ we have that
$|\tilde{\psi}\rangle = B_{i,j}|\psi\rangle$ is also a vector in
$\mathcal{H}_{c}$.

\begin{figure}[t!]
 \begin{center}
\includegraphics*[width=0.45\textwidth]{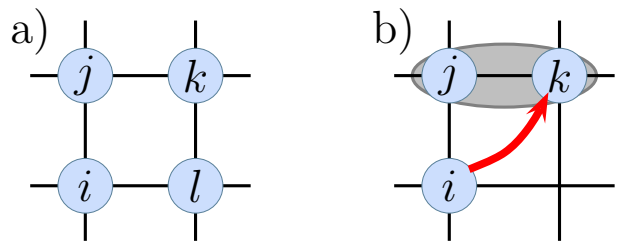}
\end{center}
\caption{ \label{fig:index} a) Indexes corresponding to each square
plaquette in the Hamiltonians $H_{J}$ and $H_{V}$.  b) Indexes
corresponding to each hopping process in $H_{t}$.  }
\end{figure}
One can easily notice that the operator $B_{i,j}$ is invariant with
respect to the gauge transformation
\ba 
a_{j} & \rightarrow & e^{i\xi_{j}}a_{j}\\ b_{j,k} & \rightarrow &
e^{i(\xi_{j}+\xi_{k})}b_{j,k}
 \ea
This $U(1)$ gauge symmetry was exploited in Ref.\onlinecite{Fradkin+Kivelson} to represent
 doped QDM as theories in which a matter field is coupled to a gauge field.
 More importantly, one can check that
{\it within the subspace} $\mathcal{H}_{c}$, the set of operators 
$\{B_{i,j} \}$ form a closed algebra similar to the one of $\{ b_{i,j} \}$.
%
%
The Hamiltonian can entirely be written in terms of these $B_{i,j}$
operators making its gauge invariance a trivial statement.  Its precise
form is given by $H= H_J + H_V + H_t$ with:
\ba H_{J}&=&-J\sum_{\square}\left\{ B^{\dagger}_{i,j}B^{\dagger}_{k,l}
B_{j,k}B_{l,i}+\hbox{H.c.}  \right\}\\
H_{V}&=&V\sum_{\square}\left\{ B^{\dagger}_{i,j}B^{\dagger}_{k,l}
B_{i,j}B_{k,l}
+ B^{\dagger}_{j,k}B^{\dagger}_{l,i} B_{j,k}B_{l,i} \right\}\\
H_{t}&=&-t\sum_{i}\left\{ B^{\dagger}_{i,j}B_{j,k} + \hbox{H.c} \right\}
\ea

It is evident that, within this formulation, the basic building blocks of
the model are created by $B^{\dagger}_{i,j}$ which corresponds to
composite particles of charge $2e$.  Namely, the model is completely
defined in terms of the constituent particle with charge $2e$.  This has
several important consequences.  In particular, the gauge invariance
requires that the energy spectrum
of the system on a torus, as a function of the magnetic flux $\Phi$
through the ``hole'' of the torus, is invariant under $\Phi
\rightarrow \Phi + \pi/e$.
(For early discussions on the $\pi/e$-flux periodicity
in the QDMs, see Refs.~\onlinecite{Thouless87,KRS88} and references therein.)
This periodicity corresponds to the unit
flux quantum for charge $2e$ objects.
However, this does not
necessarily mean that the physical elementary excitations of the system
have minimum charge $2e$.\cite{OS-fractionalization} 
The system can have a topological order which leads to 
\emph{fractionalization}; elementary excitations can have
fractions of the charge $2e$ of the constituent particle of
the microscopic Hamiltonian.
If the charge-$e$ holons are deconfined as a result of
fractionalization, they could condense to form an exotic
superconductor.

The apparent contradiction between the periodicity of
the energy spectrum in $\pi/e$ flux and
the expected flux quantization
in the unit of $2\pi/e$ in the condensate of charge-$e$ holons
is resolved by the existence of the topological
vortex excitation called vison.
Insertion of the $\pi/e$ flux corresponds to 
trapping of a vison.
Although the flux periodicity of the groundstate energy
does not distinguish an exotic charge-$e$ condensate
from the usual superconductor,
an experimental detection scheme
of the charge-$e$ condensation,
based on a ``vortex memory effect'', was proposed.\cite{SenthilFisher}
An actual experiment\cite{Bonn-vortex}
on the high-$T_c$ superconductor did not
find such a signature of charge-$e$ condensation.
Nevertheless, the exotic supercondutivity due to condensation
of charge-$e$ objects is possible in principle, and is an
interesting subject to pursue theoretically and experimentally.
Later in this paper, we will introduce a quantity which
detects a charge-$e$ condensation, and study it numerically
in several QDMs.

\subsection{Statistical transmutation symmetry}

The main advantage of the formulation in terms of composite operators
presented above is that one can prove an equivalence between a
Hamiltonian where holes are hard core bosons and another one where the
holes are fermions.  Let us consider a QDM hamiltonian with bosonic
holes, where their creation and annihilation operators $a^{\dagger}_{i}$
and $a_{j}$ satisfy bosonic commutation relations. Let us also consider
another QDM hamiltonian with fermionic holes, created and annihilated by
the set of operators $f^{\dagger}_{i}$ and $f_{j}$ which now satisfy
fermionic anti-commutation relations.  We then build the respective
composite operators:
\ba B_{i,j}&=&b_{i,j}a^{\dagger}_{i}a^{\dagger}_{j}\\
\tilde{B}_{i,j}&=&b_{i,j}f^{\dagger}_{i}f^{\dagger}_{j} \ea
As for the operators $\tilde{B}$, all quantities in the rest of the paper
with a tilde correspond to operators and coupling constants of the {\it fermionic}
representation for the holes. 
Before proceeding, there is an important statement to make. Again, one
can show that within the subspace $\mathcal{H}_{c}$, both set of
operators $\{ B_{i,j} \}$ and $\{ \tilde{B}_{i,j} \}$ form the {\it
same} closed algebra of {\it bosonic} dimer operators.  Another
important point is that the definition of the composite operators in
terms of fermions is more subtle because it is necessary to take a
prescription for the orientation of the dimers (which determines the
order of the fermions in the endpoints of each dimer).

In the following we will call {\it ``even prescription''} of a given
 plaquette an ordering prescription for the bonds such that all the
 bonds are oriented in a clockwise direction or an even number of bonds
 are oriented anti-clockwise.
By contrast we call {\it ``odd prescription''} the prescriptions
obtained from the clockwise ordering by flipping an odd number of bonds.

Notice that, since the resonance plaquettes containing $N$ dimers have
necessary $2N$ bonds, the anti-clockwise prescription(where all bonds
are oriented anti-clockwise) is always an {\it even prescription}.

\vspace{1cm}

\paragraph{Theorem:}
{ \it Given a resonant plaquette of arbitrary length with an even
prescription for the bonds, then, for the kinetic term of the dimers in
the plaquette, we have the equivalence: $H_{J}(J,bosons) \leftrightarrow
H_{J}(-J, fermions)$.  In other words, the resonance term of dimers in
the plaquette is invariant under a simultaneous change of ``statistics''
of the holes in the system (i.e. bosonic into fermionic or vice-versa)
and the sign of the dimer resonance loop amplitude $J$.  }

\vspace{0.5cm}

\begin{figure*}[t!]
 \begin{center}
\includegraphics*[width=0.99\textwidth]{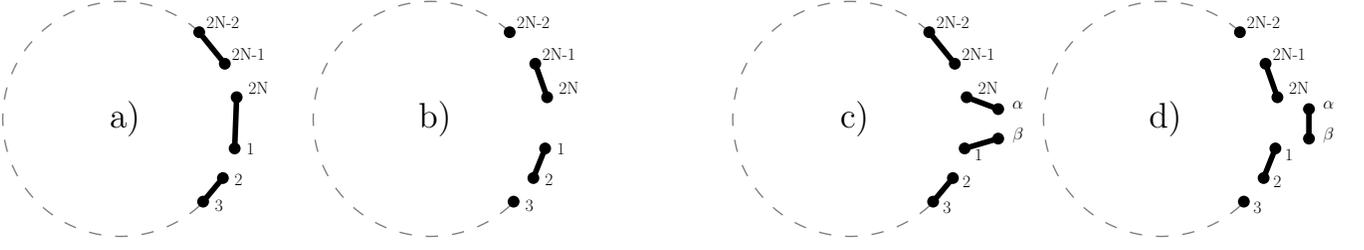}
\end{center}
\caption{ \label{fig:loop} Elements used to prove the inductive step. a)
and b) correspond to the two possible dimerization in a $N$ dimers
plaquette whereas c) and d) correspond to the plaquette with $N+1$
dimers.  }
\end{figure*}

Proof:

Consider a resonance loop containing $2N$ sites ($N$ dimers) numbered
 from 1 to $2N$ in the clockwise direction as in Fig. \ref{fig:loop}-a
 and Fig. \ref{fig:loop}-b.  The kinetic Hamiltonian for dimers
 belonging this loop can be written in terms of bond operators $b_{i,j}$
 as
\ba H^{N}_{J}&=&J
\big[b^{\dagger}_{1,2}b^{\dagger}_{3,4}b^{\dagger}_{5,6} \cdots
b^{\dagger}_{2N-3,2N-3}b^{\dagger}_{2N-1,2N} \big] \\
&\times&\big[b_{2,3}b_{4,5} \cdots b_{2N-2,2N-1} {
b_{2N,1}}\big]+\hbox{H.c}
\ea
where, the index $N$ indicates the number of dimers in the loop.

Now, we add one dimer (2 sites) to the loop, obtaining a resonance loop
 with $N+1$ dimers.  In this case, the Hamiltonian can be written as
\ba \nn H^{N+1}_{J}&=&J
\big[b^{\dagger}_{1,2}b^{\dagger}_{3,4}b^{\dagger}_{5,6} \cdots
 b^{\dagger}_{2N-3,2N-2} b^{\dagger}_{2N-1,2N} {
 b^{\dagger}_{\alpha,\beta}}\big]\\\nn &\times&\big[b_{2,3}b_{4,5}
 \cdots b_{2N-2,2N-1} { b_{2N,\alpha}b_{\beta,1} }\big]+\hbox{H.c}
\ea
Since all the operators are acting on different bonds they all commute
 and we can rearrange them in the following way
\ba \nn H^{N}_{J}&=&J
\big[ b^{\dagger}_{1,2} b^{\dagger}_{3,4} \cdots
  b^{\dagger}_{2N-3,2N-3}b^{\dagger}_{2N-1,2N}\big]\\\nn
  &\times&\big[b_{2,3}b_{4,5} \cdots b_{2N-2,2N-1} \big] {
  b_{2N,1}}+\hbox{H.c}
\ea
\ba \nn H^{N+1}_{J}&=&J
\big[b^{\dagger}_{1,2}b^{\dagger}_{3,4} \cdots b^{\dagger}_{2N-3,2N-3}
 b^{\dagger}_{2N-1,2N} \big]\\\nn &\times&\big[b_{2,3}b_{4,5} \cdots
 b_{2N-2,2N-1} \big] { b^{\dagger}_{\alpha,\beta}
 b_{2N,\alpha}b_{\beta,1} }+\hbox{H.c}
\ea
or in a compact notation
\ba
\nn
\lefteqn{H^{N}_{J}=} \\
&& J
\left( \prod_{j=1}^{N} b^{\dagger}_{2j-1,2j} \right) \left(
\prod_{j=1}^{N-1} b_{2j,2j+1} \right) { b_{2N,1}}+\hbox{H.c.}
\nn
\ea
\ba \nn H^{N+1}_{J}=J
\left( \prod_{j=1}^{N} b^{\dagger}_{2j-1,2j} \right) \left(
\prod_{j=1}^{N-1} b_{2j,2j+1} \right) { b^{\dagger}_{\alpha,\beta}
b_{2N,\alpha}b_{\beta,1} }+\hbox{H.c}
\ea
Now, we insert on the right of the Hamiltonian the string of operator
 $S^{f}_{N}=\prod_{i=1}^{2N} f_{i}f_{i}^{\dagger}$, where the index $i$
 correspond to the sites on the resonance loop.  This operator acts as
 the identity operator on the sites belonging the loop because
 $f_{i}f_{i}^{\dagger}=1$ in the absence of holes.  We start with the
 Hamiltonian $H^{N}_{J}$
\ba \nn H^{N}_{J}&=&H^{N}_{J}S^{f}_{N}\\\nn &=&J
\left( \prod_{j=1}^{N} b^{\dagger}_{2j-1,2j} \right) \left(
\prod_{j=1}^{N-1} b_{2j,2j+1} \right) { b_{2N,1}}\\\nn &\times&
f_{1}f_{1}^{\dagger}\cdots f_{2N}f_{2N}^{\dagger} +\hbox{H.c}
\ea
Now, we move the fermions to the left in order to form the composite
operators $\tilde{B}_{i,j}=b_{i,j}f^{\dagger}_{i}f^{\dagger}_{j}$
corresponding to the dimer operators inside the brackets.  Commutation
of the fermions gives a global sign.
%
\ba \nn H^{N}_{J}&=&(-1)^{\xi}J
\left( \prod_{j=1}^{N} \tilde{B}^{\dagger}_{2j-1,2j} \right) \left(
\prod_{j=1}^{N-1} \tilde{B}_{2j,2j+1} \right)\\\nn &\times& { b_{2N,1}}
f_{1}^{\dagger}f_{2N}^{\dagger} +\hbox{H.c}
\ea
We can follow exactly the same procedure in the loop with $N+1$ dimers,
 the global sign resulting from the commutation of fermion operators to
 write the products 
in terms of composite particles is the same
 that in the $N$ dimers case.  We can write for the $N+1$ case
\ba \nn H^{N+1}_{J}&=&(-1)^{\xi}J
\left( \prod_{j=1}^{N} \tilde{B}^{\dagger}_{2j-1,2j} \right) \left(
\prod_{j=1}^{N-1} \tilde{B}_{2j,2j+1} \right)\\\nn &\times&{
b^{\dagger}_{\alpha,\beta}b_{2N,\alpha}b_{\beta,1}}
f_{1}^{\dagger}f_{2N}^{\dagger}f_{\alpha}f_{\alpha}^{\dagger}f_{\beta}f_{\beta}^{\dagger}
+\hbox{H.c}
\ea
Now we can determine the change in the sign of $J$ when a dimer is added
in the loop. First we commute the operators $f_{1}^{\dagger}$ and
$f_{2N}^{\dagger}$ in $H^{N}_{J}$ to form the operator
$\tilde{B}_{2N,1}=b_{2N,j1}f^{\dagger}_{2N}f^{\dagger}_{1}$. This
commutation gives another sign to complete the global phase in the
Hamiltonian. Then we can write for the $N$ dimers case
\ba \nn H^{N}_{J}=(-1)^{\xi+1}J\!
\left( \prod_{j=1}^{N} \tilde{B}^{\dagger}_{2j-1,2j} \right)\!  \left(
\prod_{j=1}^{N-1} \tilde{B}_{2j,2j+1} \right)\!  \tilde{B}_{2N,1}
\!+\!\hbox{H.c}
\ea

In the Hamiltonian corresponding to $N+1$ dimers we have to define 3
 composite operators.  We have that
\begin{widetext}
\ba b^{\dagger}_{\alpha,\beta}b_{2N,\alpha}b_{\beta,1}
 f_{1}^{\dagger}f_{2N}^{\dagger}f_{\alpha}f_{\alpha}^{\dagger}f_{\beta}
 f_{\beta}^{\dagger} &=&
 \left(b^{\dagger}_{\alpha,\beta}f_{\beta}f_{\alpha}\right)
 b_{2N,\alpha}b_{\beta,1}
 f_{1}^{\dagger}f_{2N}^{\dagger}f_{\alpha}^{\dagger}
 f_{\beta}^{\dagger}\\ &=&
 \left(b^{\dagger}_{\alpha,\beta}f_{\beta}f_{\alpha}\right)
 \left(b_{2N,\alpha}f_{2N}^{\dagger}f_{\alpha}^{\dagger}\right)
 b_{\beta,1} f_{1}^{\dagger} f_{\beta}^{\dagger}\\ &=&
 (-1)\left(b^{\dagger}_{\alpha,\beta}f_{\beta}f_{\alpha}\right)
 \left(b_{2N,\alpha}f_{2N}^{\dagger}f_{\alpha}^{\dagger}\right)
 \left(b_{\beta,1}f_{\beta}^{\dagger}f_{1}^{\dagger}\right)\\ &=&
 (-1)\tilde{B}^{\dagger}_{\alpha,\beta} \tilde{B}_{2N,\alpha}
 \tilde{B}_{\beta,1}
\ea
\end{widetext}
Finally, the Hamiltonian corresponding to $N+1$ dimers reads
\ba \nn H^{N+1}_{J}&=&(-1)^{\xi+1}J
\left( \prod_{j=1}^{N} \tilde{B}^{\dagger}_{2j-1,2j} \right) \left(
\prod_{j=1}^{N-1} \tilde{B}_{2j,2j+1} \right)\\\nn
&\times&\tilde{B}^{\dagger}_{\alpha,\beta}\tilde{B}_{2N,\alpha}\tilde{B}_{\beta,1}
+\hbox{H.c}
%
\ea

We have proved by induction that, if for a $N$ dimers loop, the kinetic
 term acquires a given sign when the Hamiltonian is written in terms of
 fermionic composite particles, then the kinetic amplitude corresponding
 to $N+1$ dimers acquires the same sign.  To complete this mathematical
 induction proof we need check that the statement holds for the lowest
 value of $N$. The smallest possible resonance loop is given by a loop
 with only 2 dimers. It is easy to check that in this case
\ba \nn H^{2}_{J}&=& J b^{\dagger}_{1,2}b^{\dagger}_{3,4}b_{2,3}b_{4,1}
+\hbox{H.c}\\\nn
&=& J b^{\dagger}_{1,2}b^{\dagger}_{3,4}b_{2,3}b_{4,1} f_{1}
f^{\dagger}_{1} f_{2}f^{\dagger}_{2} f_{3}f^{\dagger}_{3}
f_{4}f^{\dagger}_{4} +\hbox{H.c}\\\nn
&=& (-1) J \tilde{B}^{\dagger}_{1,2}\tilde{B}^{\dagger}_{3,4}
\tilde{B}_{2,3}\tilde{B}_{4,1} +\hbox{H.c} \ea
Then we have proved that the kinetic term for a resonance loop of
arbitrary length oriented in a clockwise direction, the amplitude $J$ in
the Hamiltonian written using dimer operators $b_{i,j}$ changes to $-J$
when we write the Hamiltonian in terms of fermionic composite operators
$\tilde{B}_{i,j}$.  A trivial verification shows that the amplitude
remains unchanged when we write the Hamiltonian in terms of bosonic
operators $B_{i,j}=b_{i,j}a^{\dagger}_{i}a^{\dagger}_{j}$.
%

The result above can be re-written in the more appealing way:
 \ba H_{J}(J,\tilde{B})\equiv H_{J}(-J,B)
\ea

The proof can easily be extended to the potential term $H_{V}$. In this
case is easy to see that the bosonic and fermionic versions give the
same sign in the amplitude $V$.  \ba H_{V}(V,\tilde{B})\equiv H_{V}(V,B)
\ea

The equivalence proved above is valid for any even prescription on the
plaquette. Starting from the clockwise prescription where the results
above has been proved, if we flip two bonds this induce the commutation
of two fermionic operators and the sign remains unchanged.  But if we
flip an odd number of loops we must to commute an odd number of extra
fermionic commutations in order to form the composite operators. These
permutations gives an extra sign in the Hamiltonian.  Then is easy to
prove the following corollary:

\vspace{0.5cm}

{\it Corollary 1:}\\ { \it Given a resonant plaquette of arbitrary
length with an odd prescription for the bonds, then, for the kinetic
energy of the dimers in the plaquette, we have the equivalence:
$H_{J}(J,bosons) \leftrightarrow H_{J}(J, fermions)$.  \hfill
 }

\vspace{0.5cm}

The equivalence in the potential term do not change if we take an odd or
even prescription.  Using this property of the potential term and the
Corollary 1 we can derive the following corollary.

\vspace{0.5cm}

{\it Corollary 2:}\\
{ \it Note that we have actually proved that, if in a given lattice we
can take an even prescription for all the plaquettes involved in the
Hamiltonian. Then the equivalence \ba H(J,V, bosons)\equiv
H(-J,V, fermions) \ea
is valid for the Hamiltonian in the complete lattice.
whereas if we can take an odd prescription for all the plaquettes in the
 Hamiltonian.  We have the equivalence \ba H(J,V, bosons)\equiv
 H(J,V, fermions) \ea
\hfill 
 }

In order to complete the panorama for the doped QDM we study the
fermionic and bosonic representation of the Hamiltonian $H_t$
corresponding to the hopping of holes.

\begin{figure*}[t!]

\includegraphics*[width=0.75\textwidth]{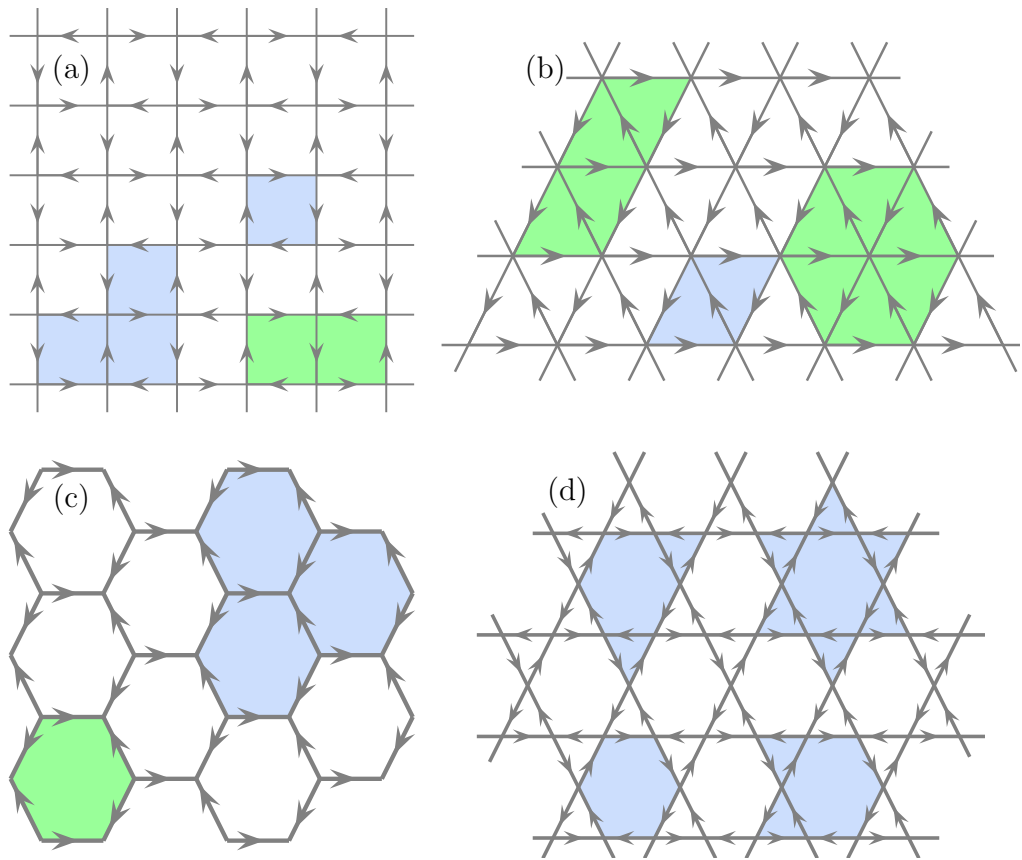}
%
%
\caption{ (Color online) Bond prescriptions on the square (a), triangular (b),
 honeycomb (c) and kagome (d) lattices.  Light-blue plaquettes have an even
 prescription while the green plaquette has an odd prescription. 
}

\label{fig:pres_square}
\end{figure*}

Consider 3 nearest-neighbors sites of the lattice as in
Fig. \ref{fig:index}-b.  In terms of bosonic holes and dimer operators
we can write a general hoping term as
\ba h^{(t)}_{i,j,k}=b^{\dagger}_{i,j}b_{j,k}a^{\dagger}_{k}a_{i} \ea
if there is no hole in the intermediate site $j$ we can add on the right
the identity as $a_{j}a_{j}^{\dagger}=1$. We then obtain
\ba h^{(t)}_{i,j,k}&=&-t \,
b^{\dagger}_{i,j}b_{j,k}a^{\dagger}_{k}a_{i}a_{j}a_{j}^{\dagger}\\ &=&-t
\, \left(b^{\dagger}_{i,j} a_{j}a_{i}\right)
\left(b_{j,k}a_{j}^{\dagger}a^{\dagger}_{k}\right) \\ &=&-t \,
B^{\dagger}_{i,j}B_{j,k} \ea
In the bosonic case we don't need to worry about the prescription in the
lattice but it is important when we study the fermionic description. In
this case we take the prescription $i\rightarrow j \rightarrow
k$. Starting from the Hamiltonian
\ba \tilde{h}^{(t)}_{i,j,k}=b^{\dagger}_{i,j}b_{j,k}f^{\dagger}_{k}f_{i}
\ea
we insert on the right the operators $f_{j}f_{j}^{\dagger}=1$
\ba \tilde{h}^{(t)}_{i,j,k}&=&-t \,
b^{\dagger}_{i,j}b_{j,k}f^{\dagger}_{k}f_{i}f_{j}f_{j}^{\dagger}\\ &=&-t
\, \left(b^{\dagger}_{i,j} f_{j}f_{i}\right)
\left(b_{j,k}f_{j}^{\dagger}f^{\dagger}_{k}\right) \\ &=&-t \,
\tilde{B}^{\dagger}_{i,j}\tilde{B}_{j,k} \ea
Using the prescription $i\rightarrow j \rightarrow k$, the
amplitude in the hopping term for the holes is the same if we use the
fermionic or bosonic versions of the composite operators.  Flipping two
arrows we have the prescription $k\rightarrow j \rightarrow i$. It is a
simple matter to see that with this prescription the hopping amplitude
is also the same for the two cases.

Then, the hopping of the holes written in terms of bosonic and fermionic
composite operators have the same amplitude $t$, provided that we use
one of the two prescriptions satisfying that the intermediate site has
one incoming and one out-coming arrow.  If we take this $i\rightarrow j
\rightarrow k$ prescription in all the sites of the lattice the arrows
follow a sort of Kirchhoff's first rule: see Fig. \ref{fig:pres_square}- (a), (b), (d).
We will call this kind of prescriptions as ``zero-current''
prescriptions.  Of course it is only possible to satisfy this
prescription in all the sites if the coordination number of the lattice
is even. An example where this is not possible is the Honeycomb lattice
(with $z=3$). In this lattice it turns out that it is not possible to
take a prescription with the same number of incoming and out-coming
arrows in each site.  See Fig.  \ref{fig:pres_square}-(c).

\section{QDM classification for different lattices}
\label{sec.classification}

\subsection{On the choice of the bond orientation prescription}

\begin{figure}[t!]
\includegraphics[width=0.45\textwidth]{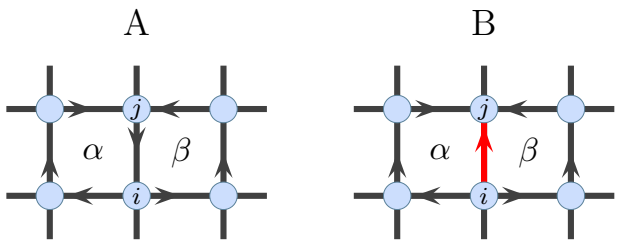}
\caption{ \label{fig:arrow_reversed} The change of prescription
corresponding to reversing the orientation of one single bond ($ij$ in
the figure) corresponds to a gauge transformation where only
configurations containing a dimer in the $ij$ bond have their sign
changed. This in turn has the effect of reversing the sign of the
flipping and hoping terms containing the bond $ij$, as for example the
flipping of plaquettes $\alpha$ and $\beta$. }
\end{figure}

As we saw in the last section, in order to prove the equivalence between
 Hamiltonians built with bosonic and fermionic operators, one needs a
 bond orientation prescription for the fermionic case.  Of course, this
 prescription is totally arbitrary and before proceeding it is important
 to clarify the issue of a different choice of prescription.  Let us
 imagine a generic lattice for which we have chosen two different
 prescriptions, A and B.  To clarify the ideas, imagine that the
 orientation of all the bonds in prescription B are the same that in
 prescription A, except for one single bond, which is connecting points
 $i$ and $j$. Then, starting from a bosonic Hamiltonian, by doing the
 transmutation, we end up with two different Hamiltonians $H_{A}$ and
 $H_{B}$ which have the same signs for all the flipping and hopping terms
 except for the ones that contain the bond $ij$.  Let us illustrate this
 with the following example: consider the square lattice in which
 prescription A is the one given in Fig.~\ref{fig:pres_square}.  Then,
 imagine a prescription B where only the arrow between sites $i$ and $j$
 is reversed, as shown in Fig. \ref{fig:arrow_reversed}.  Starting from
 the same bosonic Hamiltonian, after the statistical transmutation, we
 get the Hamiltonians $H_{A}$ and $H_{B}$.  What is the difference
 between $H_{A}$ and $H_{B}$?  They have the same signs for all the
 flipping terms, except the the ones of plaquettes $\alpha$ and $\beta$,
 which are the only two containing this reversed bond. Also, all hoping
 terms are the same except those containing the link $ij$.

Although one could naively think that these two resulting Hamiltonians
 are not equivalent, in fact they are, as can be easily seen by
 performing the following gauge transformation:
\ba \nonumber b^{\dagger}_{n,m}&\rightarrow& -b^{\dagger}_{n,m} \hbox{
if } n=i,m=j \hbox{ or } n=j,m=i\\ \nonumber
b^{\dagger}_{n,m}&\rightarrow& b^{\dagger}_{n,m} \hbox{ else. } \ea
Most generally, it is easy to convince oneself that different choices of
prescriptions give rise to apparently different Hamiltonian which in
fact are equivalent under a certain gauge transformation.  We are now
going to consider each lattice in detail and justify for each of them
the choice of prescription we have made.

\subsection{Square lattice}

For the square lattice, we consider the prescription given in
Fig. \ref{fig:pres_square}-(a). Using this prescription, the hopping
amplitude ($t$) remains equal for the bosonic and fermionic
representation $B$ and $\tilde{B}$.  On the other hand, the kinetic
amplitude corresponding to dimers ($J_{\alpha}$) changes its sign if an
even prescription is induced in the plaquette of length $\alpha$. In
Fig.  \ref{fig:pres_square} the plaquettes of lowest order are
shown. Light-blue areas correspond to even prescriptions induced in the
plaquettes while green areas correspond to odd prescriptions.  The
relative sign between the couplings in the fermionic and bosonic representations
corresponding to the 8 smallest plaquettes are presented in table
\ref{tab:square_plaq}

\begin{table}[h!]
\begin{center}
\begin{tabular}{|c|c|c|}
\hline $N$ & Loop & $\frac{\tilde{J}/\tilde{t}}{J/t}$ \\\hline
2 & $\fifi[0.3]{0.0}{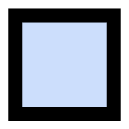}$& -1 \\\hline
3 & $\fifi[0.3]{0.0}{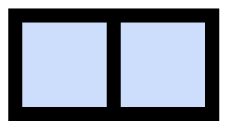}$& \, 1 \\\hline
\multirow{3}{*}{4} & $\fifi[0.3]{0.0}{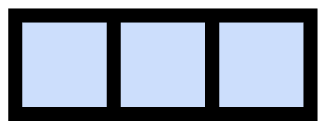}$ & -1 \\
                   \cline{2-3} & $\fifi[0.3]{-1.0}{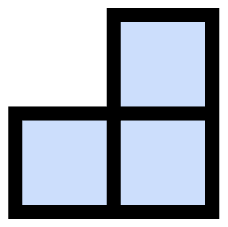}$ & -1
                   \\ \cline{2-3} & $\fifi[0.3]{-1.0}{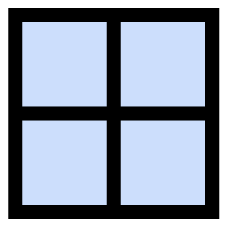}$ &
                   -1 \\\hline
\multirow{3}{*}{5} & $\fifi[0.3]{0.0}{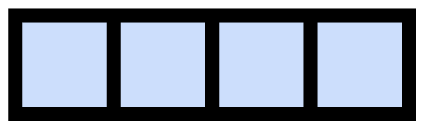}$ & \, 1 \\
                   \cline{2-3} & $\fifi[0.3]{-1.0}{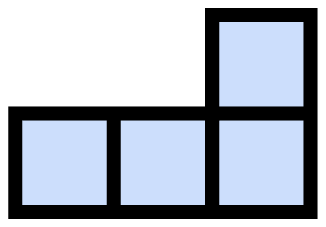}$ & \,
                   1 \\ \cline{2-3} & $\fifi[0.3]{-1.0}{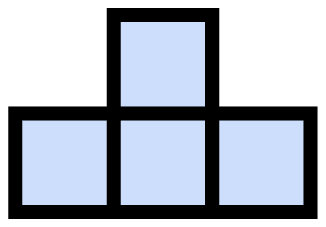}$
                   & \, 1 \\\hline
\end{tabular}
\end{center}
\caption{Values of $\tilde{J}_{\alpha}/J_{\alpha}$ corresponding to the
lowest orders of the resonant plaquettes on the square lattice. }
\label{tab:square_plaq}
\end{table}

\subsection{Triangular lattice}

For the triangular lattice, as in the square case, the coordination
 number is even and we can take a ``zero-current'' prescription as shown
 in Fig.  \ref{fig:pres_square}- (b). Then the hopping amplitudes for bosonic
 and fermionic holons have the same sign. Again, the change in the sign
 when we change from a bosonic representation of the holes to a
 fermionic one is determined by the parity of each flipping term. In
 Table \ref{tab:triangular_plaq} we show the results for flipping loops
 containing up to three dimers.

\begin{table}[h!]
\begin{center}
\begin{tabular}{|c|c|c|}
\hline $N$ & Loop & $\frac{\tilde{J}/\tilde{t}}{J/t}$ \\\hline

\multirow{3}{*}{2} & $\fifi[0.32]{0.0}{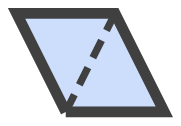}$ & -1 \\
                   \cline{2-3} & $\fifi[0.32]{0.0}{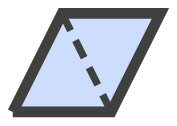}$ & -1
                   \\ \cline{2-3} & $\fifi[0.32]{-1.0}{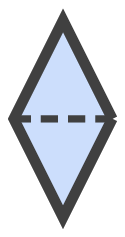}$ &
                   -1 \\\hline
\multirow{4}{*}{3} & $\fifi[0.3]{-1.0}{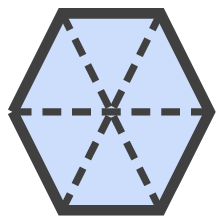}$ & \, 1 \\
                   \cline{2-3} & $\fifi[0.3]{-1.0}{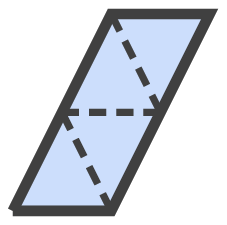}$ & \, 1
                   \\ \cline{2-3} & $\fifi[0.3]{-1.0}{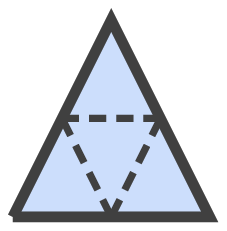}$ &
                   -1 \\\hline
\end{tabular}
\end{center}
\caption{Values of $\tilde{J}_{\alpha}/J_{\alpha}$ corresponding to the
lowest orders of the resonant plaquettes on the triangular lattice. }
\label{tab:triangular_plaq}
\end{table}

\subsection{Honeycomb lattice}

The case of the Honeycomb lattice is more subtle. The coordination
number in this lattice is $z=3$ and it is not possible to take a
``zero-current'' prescription. Therefore, it is not possible to find a
prescription in which all the hopping terms would remain the same after
the transmutation.  We then use the prescription of
Fig. \ref{fig:pres_square}-(c) in which all the hopping amplitudes for
the holes change the sign when we change to the fermionic representation
of the operators ($\tilde{t}_{n}=-t_{n}$).

The relative signs between the ratios $\tilde{J}/\tilde{t}$ and $J/t$
are presented in table (\ref{tab:honeycomb_plaq}) for plaquettes of
3, 5 and 6 dimers.

\begin{table}[h!]
\begin{center}
\begin{tabular}{|c|c|c|}
\hline $N$ & Loop & $\frac{\tilde{J}/\tilde{t}}{J/t}$ \\\hline 3 &
$\fifi[0.32]{-1.2}{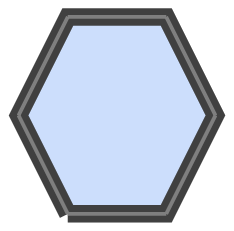}$ & -1 \\\hline
5 & $\fifi[0.32]{-2.2}{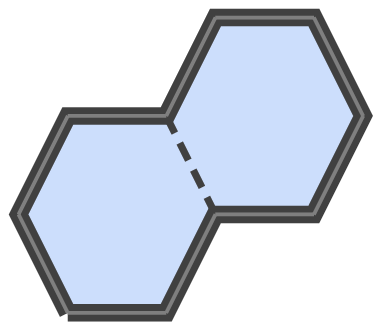}$ & -1 \\\hline
6 & $\fifi[0.32]{-3.0}{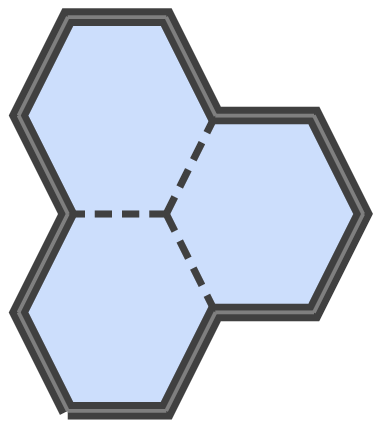}$ & \, 1 \\\hline
\end{tabular}
\end{center}
\caption{Values of $\tilde{J}_{\alpha}/J_{\alpha}$ corresponding to the
lowest orders of the resonant plaquettes on the honeycomb lattice. }
\label{tab:honeycomb_plaq}
\end{table}

\subsection{Kagome lattice}

For the kagome lattice, we have chosen the prescription depicted in figure
\ref{fig:pres_square}-(d).  All the possible allowed flipping terms up
to $12$ bonds are depicted in the table (\ref{tab:kagome_diag}) of the appendix.  Is it
interesting to note that these loops are all without exception
even. This feature is not specific to loops of short lengths and one can
convince oneself that all allowed flipping loops of arbitrary length are
even. More over, our prescription choice is such that the hoping terms
within one triangle remain invariant under the statistical
transmutation.  From this, one can conclude that a Hamiltonian with
flipping terms $\{J_l\}$ and bosonic holes is equivalent to a Hamiltonian
with fermionic holes and with the signs of all flipping terms reversed
$\{-J_l\}$.

One last remark one can make about the kagome lattice relies on its
intrinsic flexibility. Take any triangle of it and change the
orientation of the three bonds belonging to that triangle only. It is
easy to see that with the new prescription all the hopping terms,
including those belonging to the chosen triangle, do not change
signs. However, the flipping terms containing one (and only one) bond
belonging to that triangle will change their signs, {\it i.e.}, these
flipping terms in the transmutated Hamiltonian have the same sign as in
the bosonic model. What this means is that, in contrast to the other
lattice studied here, it is possible on the kagome lattice to build
gauge transformations which leave invariant the sign of all hoping
amplitude while changing the sign of "some" flipping terms (even
locally).

\subsection{Example of application of the ``statistical transmutation" symmetry}

To illustrate the power and extend of our results, we concentrate on a
couple of concrete examples taken on the square and triangular lattices
respectively. Let us consider the QDM defined on the square lattice with
only two and three dimer flipping terms. These terms correspond to the
first and second row of table (\ref{tab:square_plaq}). Its sibling model
can be defined on the triangular lattice by just considering the terms
with $N=2$ and only the second of the terms with $N=3$ in table
(\ref{tab:triangular_plaq}). In principle we would have 16 inequivalent
Hamiltonian on each case. However, our statistical transmutation result
tells us that the number of inequivalent Hamiltonians is
smaller. Indeed, we find only 8 inequivalent Hamiltonians for the case
of the triangular lattice, which we dubbed I$_{\sigma}$, II$_{\sigma}$,
III$_{\sigma}$ and IV$_{\sigma}$ where $\sigma = \pm$ corresponds to the
sign of the hoping term. From these 8 classes only 4 classes are
inequivalent for the case of the square lattice. The smaller number of
equivalence classes in the later case is due to to the equivalence $t
\leftrightarrow -t$ which is valid for the square lattice but not for
the triangular\cite{stat1}. The result is summarized in table
(\ref{tab:example_sqrt_triang}).

\begin{table}[h!]
\begin{center}
\begin{tabular}{|c|c|c|c|c|c|c|}
\hline Statistics & $\;\; J_{4}\;\; $ & $\;\; J_{6}\;\; $ & $ \;\;\; t
\;\;\; $ & Family & Refs ($\square$) & Refs ($\triangle$) \\\hline
bosons & + & + & + & I$_{+}$ & \onlinecite{poilblanc_2008}, \onlinecite{RBP}, \onlinecite{poilblanc_2006}, \onlinecite{RMP}&\onlinecite{stat1}, \onlinecite{RBP}, \onlinecite{RMP}\\\hline
bosons & + & + & - & I$_{-}$ &\onlinecite{poilblanc_2008}, \onlinecite{poilblanc_2006}
&\onlinecite{stat1}\\\hline
bosons & + & - & + & II$_{+}$ && \\\hline
bosons & + & - & - & II$_{-}$ && \\\hline
bosons & - & + & + & III$_{+}$ &\onlinecite{poilblanc_2008}& \onlinecite{stat1}\\\hline
bosons & - & + & - & III$_{-}$ &\onlinecite{poilblanc_2008}& \onlinecite{stat1}\\\hline
bosons & - & - & + & IV$_{+}$ && \\\hline
bosons & - & - & - & IV$_{-}$ && \\\hline
fermions & + & + & + & III$_{+}$ &\onlinecite{poilblanc_2008}&\onlinecite{stat1}\\\hline
fermions & + & + & - & III$_{-}$ &\onlinecite{poilblanc_2008}&\onlinecite{stat1}\\\hline
fermions & + & - & + & IV$_{+}$ &&\\\hline
fermions & + & - & - & IV$_{-}$ &&\\\hline
fermions & - & + & + & I$_{+}$ &\onlinecite{poilblanc_2008}&\onlinecite{stat1}\\\hline
fermions & - & + & - & I$_{-}$ &\onlinecite{poilblanc_2008}&\onlinecite{stat1}\\\hline
fermions & - & - & + & II$_{+}$ &&\\\hline
fermions & - & - & - & II$_{-}$ &&\\\hline
\end{tabular}
\end{center}
\caption{Classification for the doped QDM with resonant plaquettes of
 length 4 and 6.  For the square lattice the families I$_{+}$ and
 I$_{-}$ are equivalent (idem families II, III and IV). For the
 triangular lattice $J_{4}$ corresponds to the 3 resonant plaquettes
 corresponding to $N=2$ in Table \ref{tab:triangular_plaq} whereas
 $J_{6}$ corresponds to the second row of $N=3$ in the same table. 
 The two last columns show the references where such models have been studied (for $J_6=0$)
 on square ($\square$) and triangular ($\triangle$) lattices.}
 \label{tab:example_sqrt_triang}
\end{table}

\subsection{Transformation of  assisted terms}

As QDM have to be taken as low energy effective models of frustrated
anti-ferromagnets, it is important to see if other kind of term, apart
from those already mentioned here, arise in the effective QDM
Hamiltonians. Examples of derivation of the QDM Hamiltonian arising from
microscopic Heisenberg models can be found in
Ref.~\onlinecite{Heisenberg->square} for the square lattice with second and
third neighbors couplings and in \onlinecite{Heisenberg->kagome} for the
kagome lattice. In these QDM, appear a third kind of diagonal or
off-diagonal terms which has not been considered here, which are dubbed
assisted terms. An example of such terms is given in the last row of
table I of \onlinecite{Heisenberg->square}. They consist of diagonal or
off-diagonal terms of the kind of the $H_V$ and $H_J$ in Hamiltonian
(\ref{eq:H1}) but subject to the condition that a third dimer is
siting in another given neighboring bond.  Such kind of term can be
written as for example : \ba \left[ b^{\dagger}_{i,j}b^{\dagger}_{k,l}
b_{j,k}b_{l,i}+\hbox{H.c.}  \right] \left[b^{\dagger}_{m,n}
b_{m,n}\right] \ea whose effect is to flip two parallel dimers siting in
the plaquette $i,j,k,l$ provided that there is one dimer sitting in the
plaquette $m,n$.  By extending the arguments developed above, one can
show that under the statistical transmutation, this kind of terms
transform in the very same way as the correspond non-assisted term. For
example, the term written above would transform in the same way as the
term: \ba b^{\dagger}_{i,j}b^{\dagger}_{k,l} b_{j,k}b_{l,i}+\hbox{H.c.}
\ea This is simply due to the fact that assisted terms can be written as
the product of traditional diagonal or off-diagonal operators which we
know already how they transform and projectors which are written in
terms of dimer density operators which are invariant under the
statistical transmutation.


\section{Numerical investigation of QDM's on the triangular lattice}
\label{sec.numerical}

\subsection{Summary of phase diagrams in Ref.~\onlinecite{stat1}}

We now complement the analytical exact results with a numerical study.
 In our previous work \cite{stat1}, we concentrated on the triangular
 lattice because it is the best laboratory for using our analytic
 results on the statistical transmutation symmetry and for investigating
 doped dimer liquid phases.  Here we shall push further these studies
 but we start by a brief summary of the results obtained in
 Ref.~\onlinecite{stat1}.  Considering only flipping terms $J=J_4$ involving
 the shortest loops corresponding to $N=2$ in table
 (\ref{tab:triangular_plaq}) a topological ($\mathbb{Z}_{2}$) liquid can
 be stabilized at zero doping \cite{Moessner_prl_2001_Z2,Ralko_2005}
 (the sign of $J$ is irrelevant for $x=0$).  At finite doping, four
 non-equivalent families of Hamiltonians can be constructed depending on
 the signs of $t$ and $J$. Note that changing the bare statistics of the
 holons does not introduce a new class of Hamiltonian since this is
 equivalent to change the sign of $J$ as seen in the previous Sections.
 In other words, one can equivalently choose to work in the bosonic or
 fermionic representations.  In contrast, the {\it actual} statistics of
 the dressed excitations has to be studied numerically.  One can use
 e.g. the method developed in Ref.~\onlinecite{poilblanc_2008} which
 consists of investigating the node content of the wave functions.

\begin{figure}
 \begin{center}
\includegraphics*[angle=0,width=0.5\textwidth]{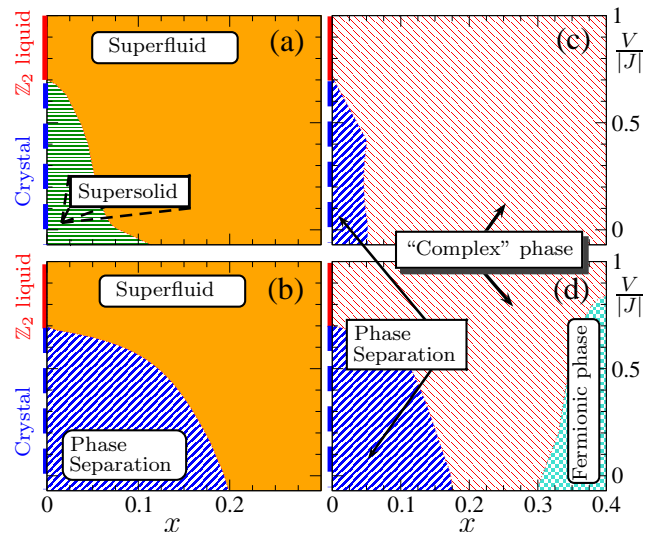}
\end{center}
\caption{ \label{fig:phasediag_triangle} (color online) Qualitative
phase diagrams of four inequivalent doped QDM's on the triangular
lattice versus doping ($x$) and $V /|J|$ for fixed $t/J=0.5$,
from Ref.~\onlinecite{stat1}.
All the
models have only flipping terms corresponding to $N=2$ in table
(\ref{tab:triangular_plaq}).  Case (a) corresponds to positive $J$ and
$t$ and bosonic holes, (b) is obtained from (a) by changing the sign of
the hoping term $t$, (c) is obtained from (a) by changing the bosons to
fermions, and (d) is obtained from (b) by changing the bosons to
fermions.}
\end{figure}

The phase diagrams of the four families of models obtained in
Ref.~\onlinecite{stat1} are reproduced in
Fig.~\ref{fig:phasediag_triangle} for convenience. At zero doping (where
the four models merge into the same $x=0$ limit) there are two
(insulating) phases, (i) a six-site cell Valence-Bond-Crystal (VBC) for $0\leq V/|J| \leq 0.7$
and a topological $\mathbb{Z}_2$ dimer-liquid above.  At finite doping,
family (a) in Fig. \ref{fig:phasediag_triangle} is the only unfrustrated
case and was studied using Green Function Monte Carlo (GFMC) methods in
Ref.~\onlinecite{RBP}.  In this model bare holons are bosonic and remain
representative of the physical excitations in the entire region of the
phase diagram, as happen also in the Perron-Frobenius square lattice
version, studied in Ref.~\onlinecite{poilblanc_2008}.  The situation is
even more interesting when $J$ is changed into $-J$, or equivalently,
bosons are changed into fermions (families (c) and (d) in
Fig. \ref{fig:phasediag_triangle}).  Finally, changing the signs of {\it
both} $J$ and $t$ of the unfrustrated model (a) leads to the complex
case (d). Note that the PS regions are further increased at $V<0$ so we
restrict here only to $V>0$.

When $J$ becomes comparable to the holon average kinetic energy (of
order $xt$) holons may be macroscopically expelled from the dimer
fluctuating background, in order to minimize the dimer resonance
energy. The question of {\it phase separation} (PS), {\it i.e.} the
possibility for the system to spontaneously undergo a macroscopic
segregation into two phases with different hole concentrations, was
considered in Ref.~\onlinecite{stat1}.  In order to perform a Maxwell
construction one can define: $s(x) = [e(x)-e(0)]/x$, where $e(x)$ is the
energy per site at doping $x=n_h/N$ ($n_h$ is the number of holons in
the system and $N$ the number of sites).  In the case of PS, the energy
presents a change of curvature at a critical doping $x_c$ corresponding
to the minimum of $s(x)$ as a function of $x$.  The fact that the local
curvature of $e(x)$ at $x = 0$ is negative then implies that the two
separated phases will have $x = x_c$ and $x = 0$ (the undoped insulator)
hole concentrations.  This study revealed that all frustrated (b-d)
models show a finite PS region (shown in blue in
Fig. \ref{fig:phasediag_triangle}(b-d)) at low doping. The extension vs
$V/|J|$ of this region seems to coincide with the $x=0$ VBC. In
contrast, the non-frustrated (a) model does not show phase separation at
$V>0$ (as already seen in GFMC simulations) but, rather, shows a
homogeneous region at finite doping where VBC order survives. Because of
the coexisting non-zero superfluid order (U(1) symmetry breaking, see
below), this phase can be viewed as a ``supersolid" (SS).  Here super
solidity involves {\it hole pairing} in the vicinity of a (insulating)
VBC phase (and in the absence of PS), as also found in the {\it
frustrated} doped QDM on the square lattice~\cite{poilblanc_2008} or in
doped frustrated spin-1/2 quantum magnets.~\cite{tJ1J2}

Another important quantity used in Ref.~\onlinecite{stat1} is the sign
operator defined in Ref.~\onlinecite{poilblanc_2008} which provides a
quantitative analysis of the nodal structure of the wave function and
hence gives insights about the statistical nature of the holons, {\it
i.e.}  whether they truly behave as bosons or fermions.  Such an
analysis clearly showed that the GS of models (a) and (b) have the same
nodal structure as a superfluid.  For models (c) and (d), we dubbed a
``complex phase" as the statistics of dressed excitations does not
correspond solely to bosons or fermions.  Family (d) also shows, in
addition, an interesting fermion reconstruction at large doping, which
we call ``fermionic phase". In this work we have extended the study
of the nodal structure for larger values of the doping than the ones
of Fig. \ref{fig:phasediag_triangle}. Our results clearly show that
for $x \gtrsim 6$ the elementary excitation behaves as bosons for the 
four models. This means in particular that in model (d) a dynamical 
statistical transmutation took place in which each Fermion bound to
a vison in order to form a boson. As we show below, this result have important
consequences in the nature of the superfluid phases that we find for 
those values of the doping. 

Finally, an Aharonov-Bohm flux can be inserted in one of the hole of the
torus, as done in Ref.~[\onlinecite{RMP}] for the doped QDM on the
square lattice.  A superfluid is characterized by well-defined minima in
the ground state energy separated by a finite barrier in the
thermodynamic limit. {\it A contrario}, a typical signature of (weakly
interacting) fermions, a flat energy profile is expected even on such a
small cluster \cite{Poilblanc1991}. Here, it was reported in
Ref.~\onlinecite{stat1} that the ground state energy has well-defined
minima quantized at half a flux quantum for all family of models at
$x\sim 0.25$.
This might appear as an evidence for condensation of 
charge-$2e$ particles.
However, as already discussed in Sec.~\ref{Comp-part},
the $\pi/e$-flux periodicity is a generic feature
of doped QDMs and does not rule out the possibility of
condensation of deconfined charge-$e$ holons.
Next, we will characterize more thoroughly the nature of the
superfluid phases, by introducing
a gauge-invariant holon Green's function to distinguish
the charge-$e$ condensation from the usual charge-$2e$ condensation.

\subsection{New correlations to explore the nature of the superfluid phases}

In order to understand the nature of the superfluid phases the effective
charge of the quasiparticles that condensate have to be determined -
either charge-$e$ or charge-$2e$ quasiparticles. This is related to the
mechanism that leads to the spontaneous breaking of the U(1) symmetry
expected in a superfluid.  As a first attempt, one could naively use the
correlation function $\langle a^{\dagger}_{k}a_{j}\rangle$, but it is
not compatible with the constraint (\ref{eq:constraint}).
In other words, it is not gauge invariant and thus this correlation function
is zero.  To satisfy the constraint, or equivalently the gauge invariance, we need to write correlations in
terms of operators $B$.  As the hopping of holes can be written in terms
of operators $B$ we can move one of the holes between two distant sites
by applying a string of $B^\dagger B$'s.

\begin{table*}[tb!]
\begin{center}
\begin{tabular}{|c|c|c|c|c|c|c|c|}
\hline $\frac{\hbox{Observables}\rightarrow}{\hbox{Phases}\downarrow}$ &
$\kappa$ & $\langle b_{i,j}^{\dagger}b_{i,j}b_{k,l}^{\dagger}b_{k,l}
\rangle $ & $\langle a_k^{\dagger} a_l^{\dagger}a_i a_j \rangle $ &
$\langle a_i^{\dagger}S_{i,j} a_j \rangle $ & Sign$_{B}$ & Sign$_{F}$ &
Flux periodicity \\\hline
PS & $< 0$ & - & - & - & - & - & - \\\hline
VBC & $> 0$ & LR & SR & SR & - & - & $2e$ \\\hline
SS & $> 0$ & LR & LR & SR & 1 & 0 & $2e$ \\\hline
$2e$-SF & $> 0$ & SR & LR & SR & $0<\hbox{Sign}_{B}<1$ &
 $0<\hbox{Sign}_{F}<1$ & $2e$ \\\hline
$e$-SF & $> 0$ & SR & LR (weak) & LR & 1 & 0 & $2e$ \\\hline
Bose-liquid & $> 0$ & SR & SR & SR & 1 & 0 & $2e$ \\\hline
Fermi-liquid & $> 0$ & SR & SR & SR & 0 & 1 & $2e$ \\\hline
``Complex" phase & $> 0$ & SR & SR & SR & $0<\hbox{Sign}_{B}<1$ &
 $0<\hbox{Sign}_{F}<1$ & $2e$ \\\hline
\end{tabular}
\end{center}
\caption{Classification of the possible phases, including various superfluid (SF) phases, that may occur in doped
QDM's on the triangular lattice.  Such phases can be distinguished from
the sign of the compressibility $\kappa$, the long-distance properties
(``SR" means short-range, ``LR" means long-range) of various
correlations, or the effective charge deduced from periodicity of the GS
energy versus a magnetic flux inserted through a
torus. $\hbox{Sign}_{B}$ and $\hbox{Sign}_{F}$ were defined in
Ref.~\onlinecite{poilblanc_2008} to analyze the node content of the GS
wave function.}  \label{tab:phase_classification}
\end{table*}

\paragraph{Gauge invariant holon Green's function -}

In the subspace where the constraint is satisfied pairs of holons
operators $a_{i}a^{\dagger}_{i}$ acting on sites without holes are equal
to the identity.  Then the holon Green's function
we want to calculate is given by
\ba G^{(b)}_{i,j}&=&\sum_{n}\langle a^{\dagger}_{i}\;
\mathcal{S}^{(n)}_{j,i}\; a_{j}\rangle \label{Gij_b} \ea
where $ \mathcal{S}^{(n)}_{j,i} =
b_{j,n_1}^{\dagger}b_{n_1,n_2}b_{n_2,n_3}^{\dagger}
b_{n_3,n_4}... b_{n_{N-1},n_N}^{\dagger}b_{n_N,i}$ is a string operator
between the sites $i$ and $j$ following the path $n$. The label $(b)$
indicates that the holes are taken as bosons.  Similarly, the fermionic
version of the Green's function is written as
\ba G^{(f)}_{i,j}&=&\sum_{n}\langle f^{\dagger}_{i}\;
\tilde{\mathcal{S}}^{(n)}_{j,i}\; f_{j}\rangle \label{Gij_f} \ea where $
\tilde{\mathcal{S}}_{j,i} =
\tilde{b}_{j,n_1}^{\dagger}\tilde{b}_{n_1,n_2}
\tilde{b}_{n_2,n_3}^{\dagger}\tilde{b}_{n_3,n_4}... \tilde{b}_{n_{N-1},n_N}^{\dagger}\tilde{b}_{n_N,i}$.
Note that, since there are many ways of moving a hole between two sites,
the gauge invariance alone does not uniquely determine
the definition of the holon Green's function.
Here we adopt the definition with a sum over all
possible strings (labeled by $n$) connecting the two sites
$i$ and $j$, with the same coefficient. This definition appears most natural to us, as well as in numerical implementation. We expect other definitions with some restrictions on strings would also work as an order parameter. However, the present definition looks advantageous in numerical calculations, since it can efficiently detect the holon condensation with the summation over all possible strings.

Using a complete basis of dimer/hole
configurations the correlations can be re-written as, \ba
G^{(b)}_{i,j}&=&\sum_{n}\sum_{\alpha,\beta} \langle\psi|\alpha\rangle
\langle\beta|\psi\rangle \langle \alpha | a^{\dagger}_{i}\;
\mathcal{S}^{(n)}_{j,i}\; a_{j}|\beta\rangle\\
G^{(f)}_{i,j}&=&\sum_{n}\sum_{\alpha,\beta} \langle\psi|\alpha\rangle
\langle\beta|\psi\rangle \langle \alpha | f^{\dagger}_{i}\;
\tilde{\mathcal{S}}^{(n)}_{j,i}\; f_{j}|\beta\rangle \ea

The holon Green's function can be also written in terms of the composite
operators only,
\ba
G^{(b)}_{i,j}&=&\sum_{\{n\}}\langle B^{\dagger}_{N-1,N}B_{N,i}\cdots
B^{\dagger}_{2,3}B_{3,4}B^{\dagger}_{1,j}B_{1,2}\rangle\\
G^{(f)}_{i,j}&=&\sum_{\{n\}}\langle
\tilde{B}^{\dagger}_{N-1,N}\tilde{B}_{N,i}\cdots
\tilde{B}^{\dagger}_{2,3}
\tilde{B}_{3,4}\tilde{B}^{\dagger}_{1,j}\tilde{B}_{1,2}\rangle, \ea
where the sum is over all possible paths between sites $i$ and $j$.  One
can use this representation to show that the two correlations are in
fact equal, up to an irrelevant sign. In other words, we have:
\ba G^{(b)}_{i,j}=\pm G^{(f)}_{i,j} \, ,
\ea where the relative sign depends only on the {\it relative} distance
between $i$ and $j$.  Off-diagonal long range order of $G_{i,j}$ is a
fingerprint of the spontaneous breaking of the U(1) gauge symmetry
(associated to charge conservation).  It also implies that the
condensing quasi-particles have charge $e$.  
It is also interesting to observe that at large doping, this Green function 
must necessary have an exponential decay (see below). Indeed, 
the Green function for being non zero need at least one path in which 
dimers are present all along it (the string). For a large concentration of holes $x \to 1$  
it is more and more unlikely to find a path with dimers on it so that 
the Green function $G_{i,j}$ should roughly decay as $(1-x)^L$ were $L$ is the distance
between points $i$ and $j$. 

Another observable
which can detect the charge-$e$ condensation is
\ba F_{i,j}=\langle a^{\dagger}_{i}\;
\mathcal{S}^{(n)}_{j,i}\; a^{\dagger}_{j}\rangle\, , \ea
when a symmetry-breaking groundstate (when it occurs) formed by
superposition over different dimer-number sectors is used.
Roughly speaking, this corresponds to the square of the
expectation value of a single hole creation
operator $\langle a^{\dagger} \rangle$ in the groundstate,
defined in a gauge-invariant manner.
Such an expression
is however less convenient to compute numerically
in finite-size systems and will not be used.

The scenario of Bose condensation of polarized spinons (the holons in
our current formulation) under an applied magnetic field advocated in
Ref.~\onlinecite{RBP}, implicitly implies long-range order of
$G_{ij}$. We wish here to substantiate such a scenario by an explicit
computation of this correlation function.

\paragraph{Hole pair correlations -}
If $G_{i,j}$ is short-ranged, there is no condensation of charge-$e$ holons.
However, spontaneous U(1) symmetry breaking and,
hence, superfluidity can still occur
provided the (hole) pair-pair correlation, \ba P_{i,j,k,l}=\langle
B_{i,j}B_{k,l}^\dagger\rangle\, \label{Pijkl} \ea exhibits long range
order.
The pair-pair operator is connected to the square of the single holon 
effective hopping operator in a complicated manner. 
Namely,
\begin{eqnarray}
&&(a^{\dagger}_{i}\;\sum_{n} \mathcal{S}^{(n)}_{k,i}\; a_{k})
(a^{\dagger}_{j}\;\sum_{n'} \mathcal{S}^{(n')}_{l,j}\; a_{l})\nonumber\\
&+&(a^{\dagger}_{i}\;\sum_{n} \mathcal{S}^{(n)}_{l,i}\; a_{l})
(a^{\dagger}_{j}\;\sum_{n'} \mathcal{S}^{(n')}_{k,j}\; a_{k})\nonumber\\
&=&B_{i,j}B_{k,l}^{\dagger}+ \{{\rm loop\,\, terms}\},
\end{eqnarray}
where the first part of the r.h.s. is obtained from a ``closure relation" involving
all pairs of ``retraceable" strings $n'=\bar{n}$ i.e.,
\begin{equation}
\sum_{n} (\mathcal{S}^{(n)}_{k,i}
\mathcal{S}^{(\bar{n})}_{l,j}+\mathcal{S}^{(n)}_{l,i}
\mathcal{S}^{(\bar{n})}_{k,j})=b_{ij} b_{kl}^\dagger,
\end{equation}
and the rest
corresponds to pair hopping dressed with extra loop fluctuations.
This suggests that it is physically meaningful to write the pair correlations as,
\begin{equation}
P_{ijkl} = G_{ik} G_{jl} + G_{il} G_{jk}, + P_{ijkl}^c
\label{Eq:MF}
\end{equation}
where the first two terms can be viewed as the ``mean-field" contribution and $P_{ijkl}^c$ stands for the  "connected" part in which we remove all the processes involving compositions 
of single holon hoppings. In particular, both sides scale like $x^2$ ($(1-x)^2$) when $x\rightarrow 0$ ($x\rightarrow 1$) so that it is convenient to normalize $P_{ijkl}$ ($G_{ij}$) by $x^2 (1-x)^2$ ($x(1-x)$). Eq.~(\ref{Eq:MF}) shows that LRO in the 
holon Green function $G_{ik}\rightarrow G_\infty$ characteristic of the 
charge-$e$ superfluid will induce indirectly LRO in the pair-pair correlation, $P\sim G_\infty^2$. In contrast, the conventional, charge-$2e$ superfluid is defined by LRO in the connected part {\it together with short-ranged holon Green's function}.

\paragraph{dimer-dimer correlations -}
We finish by recalling that the dimer-dimer correlations are expressed
in terms of the dimer number operators $b^{\dagger}_{i,j}b_{i,j}$ as \ba
N_{i,j,k,l}=\langle
b^{\dagger}_{i,j}b_{i,j}b^{\dagger}_{k,l}b_{k,l}\rangle\, , \ea where
sites $i$ and $j$ on one hand, and $k$ and $l$ on the other hand, are
nearest neighbor sites. Long-range order in this correlation function is
characteristic of VBC order.  The wave vector at which the associated
structure factor diverges defines the VBC wave vector.

In principle, one can use the new correlations $G_{i,j}$ and
$P_{i,j,k,l}$ to refine the previous phase diagrams ($N_{i,j,k,l}$ was
used in previous work to determine the VBC and SS regions).  To ease the
analysis of the numerical results of the doped QDM's, a classification
of the various possible phases based on simple considerations is
provided in table~\ref{tab:phase_classification}.

We note that, there is no phase where there is
a charge-$e$ condensation simultaneously with
a dimer long-range order (``LR'' for both
$\langle b_{i,j}^{\dagger}b_{i,j}b_{k,l}^{\dagger}b_{k,l}\rangle$ and
$\langle a_i^{\dagger}S_{i,j} a_j \rangle $.)
This is because existence of the dimer long-range order
leads to confinement of holons.

\subsection{Numerical results}
\begin{figure}[h!]
\begin{center}
\includegraphics*[angle=0,width=0.45\textwidth]{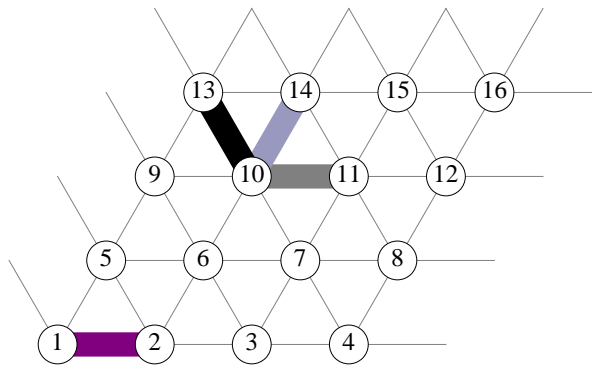}
\end{center} \caption{ \label{fig:cluster} 16 site-cluster: labeling of the
sites (numbered circles) and reference bond (purple bond) used respectively in
the definition of Green's functions and the pair correlations. The bonds are
labeled according to one of the sites connected to them and by a direction, as
shown in the example (here site 10).}
\label{Fig:cluster}
\end{figure}

\begin{figure*}[h!]
 \begin{center}
\includegraphics*[angle=0,width=0.95\textwidth]{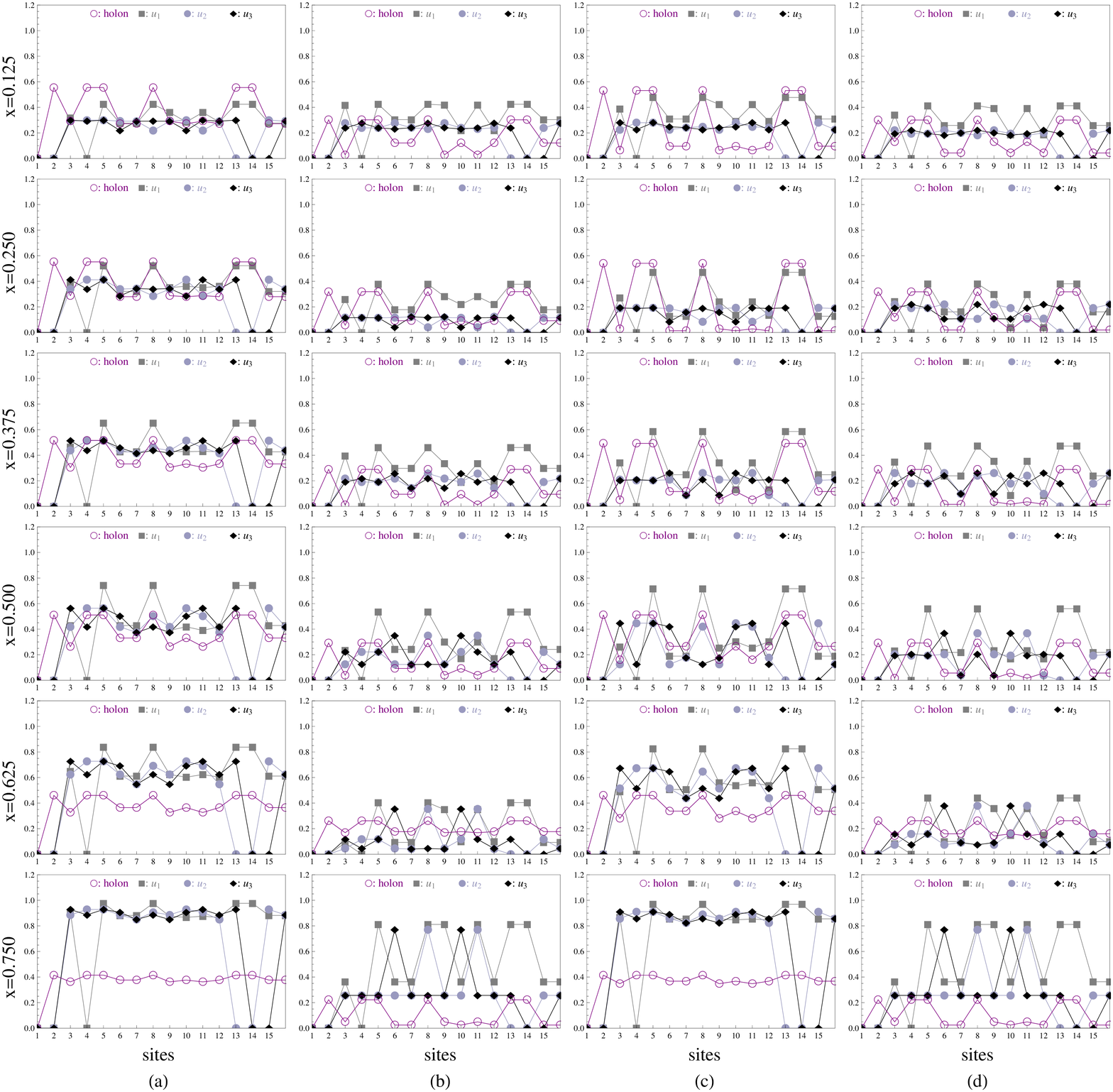}
\end{center}
\caption{ \label{fig:corr} (color online) Holon Green's function (open circles) and {\it square root} of the absolute value of the 
pair-pair correlations (filled symbols) for parameters $V=0.3$,
$|J|=1.0$ and $|t| = 0.5$, at various densities ranging from  $x=0.125$ to $x = 0.75$ (2 to 12 holons on 16 sites)
and for the 4 classes (a-d) of models defined in the caption of
Fig.~\ref{fig:phasediag_triangle}.
The pair-pair correlation $P_{i,j,k,l}$ is defined by a reference bond
orientation $\vec{\tau}_{j}-\vec{\tau}_{i}=\vec{u}_\alpha$ and the
orientation of the final bond
$\vec{\tau}_{l}-\vec{\tau}_{k}=\vec{u}_\beta$. In our case, we chose
$\vec{u}_\alpha = \vec{u}_1$ and we consider three cases for
$\vec{u}_\beta$: $\vec{u}_1$ (filled squares), $\vec{u}_2$ (filled
circles) and $\vec{u}_3$ (filled diamonds) -- see Fig.~\protect\ref{Fig:cluster}.
}
\end{figure*}

In Fig.~\ref{fig:corr} are displayed both the holon Green's 
functions (Eqs.~\eqref{Gij_b} and \eqref{Gij_f}) and the {\it square-root} of the pair-pair
correlations (Eq.~\eqref{Pijkl}) computed by numerical exact diagonalization on a 16-site triangular
cluster, varying the hole density from $x=0.125$ (low hole concentration) 
to $x=0.75$ (low dimer concentration), from top to bottom. For convenience, both the Green's functions and the 
{\it square-root} of the pair-pair correlations
are normalized by $x(1-x)$ to be able to use the same scale for all densities. The four hamiltonian 
classes defined previously (see e.g. caption of Fig.~\ref{fig:phasediag_triangle}) are
depicted in parallel panels (a), (b), (c) and (d) respectively.
For all of them. we chose the parameters $V=0.3$, $|J| = 1.0$
and $|t| = 0.5$ for which, at holon density $x=0.25$, the system is
either in a superfluid phase [Fig.~\ref{fig:phasediag_triangle}(a,b)] or
in the ``complex" phase [Fig.~\ref{fig:phasediag_triangle}(c,d)]
depending on the QDM class.

Let us first discuss the data at the lowest hole densities $x\le 0.5$.
As one can see, only model (a) presents a large amplitude of the holon Green's
function away from the reference site (the largest disk) and
its six neighbors.  While a definite conclusion cannot be drawn from
such small system, a direct comparison between model (a) and the three
others reveals a clear change of behavior. Indeed, for models (b), (c)
and (d), the holon Green's function decreases at the largest available distances to significantly smaller values, except
maybe for model (c) around $x\sim 0.5$.
On the other hand, the pair-pair correlations, for which the reference
link is in the $\vec{u}_1$ direction, show convergence with bond separation to a uniform value
for model (a) in all relative directions of the two bonds, while these correlations are
strongly reduced for directions differing from that of the reference bond in the
other models.
Hence, our data are clearly compatible with model (a) being in the $e$-superfluid phase described
in Table \ref{tab:phase_classification}, 
unambiguously revealing strong signals simultaneously in the holon Green
function and the pair-pair correlation. Note that we also checked
that the dimer-dimer correlations (not presented here) remain SR in
model (a) hence reinforcing the previous claim. The behavior of the other models at low to moderate doping
is less clear, with quite smaller amplitudes of $G_{ij}$ and $P_{ijkl}$ at the largest 
available distances. We can however recognize a possible $2e$-superfluid phase in model (c) after the phase separation zone and up to values of $x \simeq 0.2$.

While increasing holon density, from $x=0.5$ to $x=0.75$, we observe that the data for the bosonic and fermonic 
models become identical, both for $t>0$ or $t<0$. This can be understood by the fact that the (bosonic) dimers become then
the relevant entities instead of the holes. 
This implies, in particular, that statistical transmutation or pairing must occur for increasing 
$x$ for the models where fermionic statistics is expected at small $x$ (models (c) and (d)).
This is indeed confirmed by the analysis of the nodal structure of the wave functions that we performed for $x=0.625$.
In this sense the complex phase found in model (c) is probably the region in which fermions bound to visons and transmute, in order to resemble the bosonic excitations of model (a).
Note also that charge $e$ superfluidity seems to occur  at $x=0.625$ in models (a) and (c) (which seems equivalent for this doping).

\begin{figure}[t!]
\begin{center}
\includegraphics*[angle=0,width=0.45\textwidth]{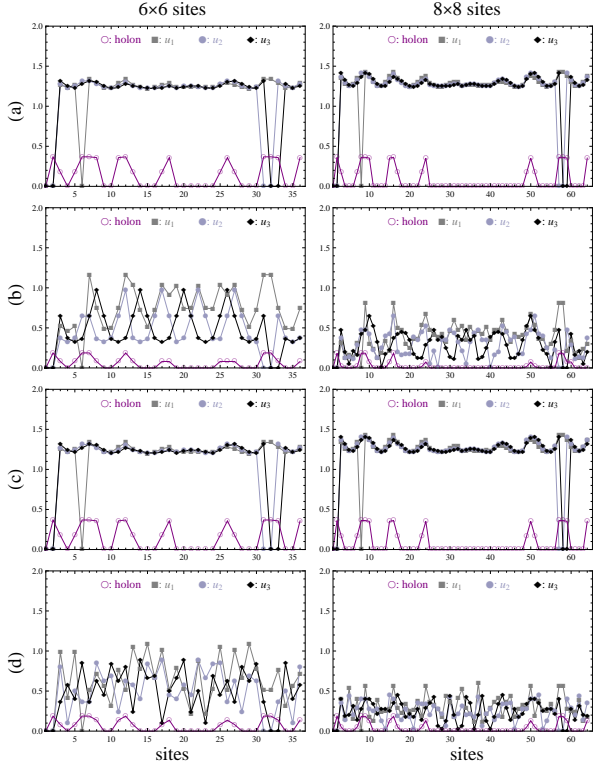}
\end{center}
\caption{ \label{fig:large_x} 
Holon Green's function (open circles) and {\it square root} of the absolute value of the 
pair-pair correlations (filled symbols) for parameters $V=0.3$,
$|J|=1.0$ and $|t| = 0.5$, at a low dimer density $1-x\simeq 0.1$
for the 4 classes (a-d) of models defined in the caption of
Fig.~\ref{fig:phasediag_triangle}. Left: 2 dimers on a 36-site cluster ($x\sim 0.89$).
Right: 3 dimers on a 64-site cluster ($x\sim 0.91$). 
}
\end{figure}

However, at larger $x$ corresponding to a dilute gas of dimers
one expects to eventually recover a charge $2e$ superfluid via a continuous (second order) or
discontinuous (first order) phase transition. This seems to occur already for $x=0.75$ for models (b) and (d),
for which only the pair-pair correlations are sizable at the largest distances.
We have checked that for a lower dimer density of $1-x\sim 0.1$ $G_{ij}$ is short range for {\it all} models as reported in
Fig.~\ref{fig:large_x}. In the limit of a very dilute gas of dimers, pairing between dimers because of the kinetic term $J$ is also a possibility.
 This could result in either phase separation or in an homogeneous phase in which both $G_{ij}$ and $P_{ijkl}$ 
are short ranged but which is nevertheless superfluid, of coherent dimer pairs of charge 4e. We have checked that
 there is not phase separation in none of the four models at those large values of $x$.
 We have also looked at the energy difference between two and one dimers and found that pairing is indeed
 favored in models (b) and (d) which may explain the drop of the $P_{ijkl}$ in Figure \ref{fig:large_x} for
 $x \gtrsim 0.9$.

\begin{figure}[t!]
\begin{center}
\includegraphics*[angle=0,width=0.5\textwidth]{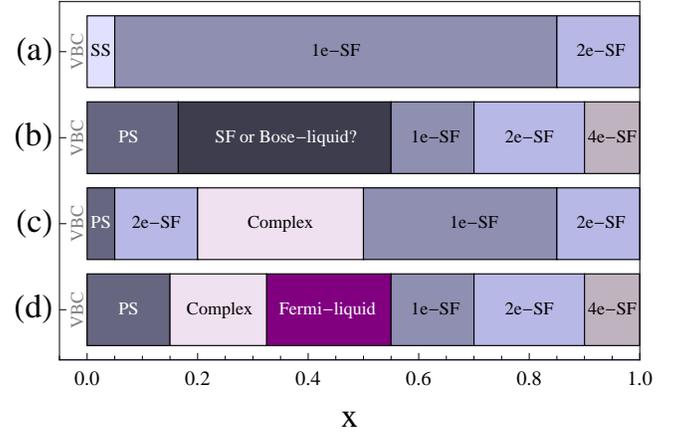}
\end{center}
\caption{ \label{fig:1dpd} Phase diagrams of the four models (a), (b), (c) and
(d) at $V/|J| = 0.3$ and $t/|J|=0.5$ derived from Fig.\ref{fig:phasediag_triangle} and
Fig.\ref{fig:corr}. All have
both the $e$-superfluid ($e$-SF) and $2e$-superfluid ($2e$-SF) phases
at different doping depending on the model.
}
\end{figure}

Based on Figs.~\ref{fig:phasediag_triangle}, \ref{fig:corr} and \ref{fig:large_x}, we have
extracted the qualitative phase diagrams for the four models at fixed $V/|J| =
0.3$ and $t/|J|=0.5$ as a function of doping $x$. They are depicted in Fig.\ref{fig:1dpd}.
Charge $e$ superfluidity seems to occur in all models, with the largest occurrence in model (a).
For intermediate doping, models (b) and (d) seem to present short range correlations for both one and two particle Green functions. This behavior suggest an uncondensed phase which in the case of model (d) would correspond to a Fermi-liquid state. Since elementary excitation in model (b) are bosonic  the presence of an uncondensed phase points toward an exotic  Bose-liquid state, although this statement should require a more detailed study (using clusters of a much bigger size) which is beyond the scope of the present article. For $x$ big, as expected, all models exhibit a $2e$-superfluid phase followed by a
charge $4e$ superfluid phase in models (b) and (d) due to dimer pairing.

\section{Connection to Bose-Hubbard models}
\label{sec.BH}

We finish this work by discussing the connection to Bose-Hubbard like
models which do not contain {\it a priori} the ice-rule
constraint. However, the physics of the doped QDM can emerge naturally
when some form of large repulsion between the itinerant bosons is
considered, hence providing emergence of fractionalized
excitations.~\cite{Bose-Hubbard_kagome}

In Ref.~\onlinecite{Bose-Hubbard_kagome} it was introduced a simple
model of hard-core bosons hopping ($t$) on a kagome lattice with a boson
repulsion $\Vhex$ favoring the smallest number of bosons in each
hexagon.
\ba H=-t \sum_{\langle i,j \rangle}
(d^{\dagger}_{i}d_{j}+d_{i}d^{\dagger}_{j})
+V_{\hexagono}\sum_{\hexagono}(n_{\hexagono})^2 \ea
where $d^{\dagger}_{i}$ creates a boson on site $i$ and
$n_{\hexagono}=\sum_{i=1}^{6}d^{\dagger}_{i}d_{i}$ is the number of
bosons in a hexagonal plaquette.  When the boson density is $\rho=1/2$ a
large $\Vhex/t$ stabilizes an insulating phase whose quantum dynamics is
described by a generalized QDM on the triangular lattice with exactly 3
dimers per site. The insulating phase is a $\mathbb{Z}_2$ topological
liquid. When $t<0$ the model is not frustrated and can be studied with
QMC~: the superfluid-insulator transition was argued to be a novel
non-conventional {\it fractional critical
point}.~\cite{Bose-Hubbard_kagome}

\begin{figure}
 \begin{center}
\includegraphics*[angle=0,width=0.45\textwidth]{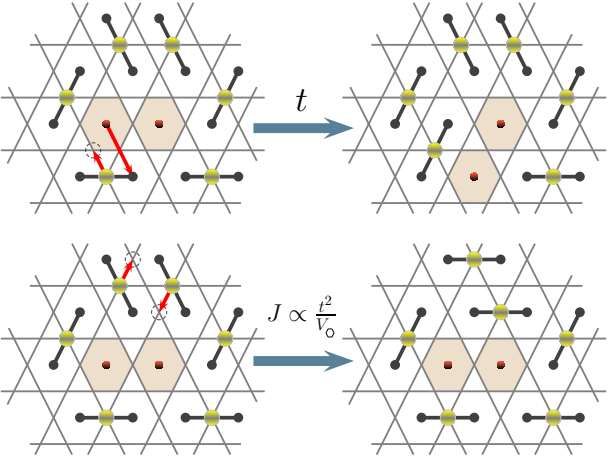}
\end{center}
\caption{ \label{fig:kagomeBH} (color online) Hard-core bosons on the kagome
lattice. Covering the lattice with one, and only one, boson per hexagonal
plaquette (ice-rule constraint), we can identify distribution of bosons with a
dimer hard-core covering on the triangular lattice. Removing a boson of charge
$2e$ creates two defect hexagons (shaded).  Each defect (hole) has charge $e$
and can move on the lattice. Coherent hoping of two hard-core bosons
corresponds to a flipping process in the dimer model. } \end{figure}

To make the connection with some of our doped QDM's we shall assume here the
microscopic $d$ bosons have charge $-2e$ and their density is set to
$\rho=\frac{1}{6}(1-x/2)$, $x<<1$. In that case, as shown in
Fig.~\ref{fig:kagomeBH}, for $\Vhex/t\rightarrow\infty$ the lowest-energy
configuration space ($E = N V_{\hexagono}  /3$) is given by all hardcore
dimer-coverings on an effective triangular lattice, where each $d$-boson has
been replaced by a dimer connecting two sites of the triangular lattice.
Such configurations respect a local {\it ice-rule} constraint with one, and
only one, boson per hexagon.
When
$x=0$, moving a single $d$-boson violates this ice-rule so one has to move at
least two simultaneously.  This process of amplitude $J=t^2/\Vhex$ corresponds
exactly to a dimer flip on a lozenge, identical to the one of the QDM. Strictly
speaking this mapping onto the QDM does not involve any dimer-dimer repulsion
$V$.  However, one can add a small  third-nearest neighbor density-density
repulsion $V_{\rm dd}<<\Vhex$ between the $d$-bosons located on different
hexagon. In the mapping for large $\Vhex/t$, this interaction
translates directly into the dimer-dimer repulsion $V=V_{\rm dd}$.
Thus, by tuning $V_{\rm dd}$ in the Bose-Hubbard model,
the topological $\mathbb{Z}_2$ (insulating) liquid can be stabilized.

When $x\ne 0$ an empty site on the original kagome lattice, corresponds
to two ``defect" hexagons carrying an overall charge $2e$ w.r.t. the
insulating GS.  It is easy to see (Fig.~\ref{fig:kagomeBH}) that each
``defect" hexagon can move independently on the effective triangular
lattice by simple processes that involve a single $d$-boson hopping.
Therefore, each ``defect" can be considered as an effective charge-$e$
hole on the triangular lattice.  The amplitude $t$ of the hole hopping
is the same as the one of the microscopic $d$-boson Hubbard model.  When
a $d$-boson of charge $-2e$ hops by a lattice spacing $a$, the effective
hole of charge $e$ hops by a distance $2a$ so that the charge center of
mass is conserved. Note also that the hole density in the effective QDM
is $x$.

\begin{figure}
 \begin{center}
\includegraphics*[angle=0,width=0.45\textwidth]{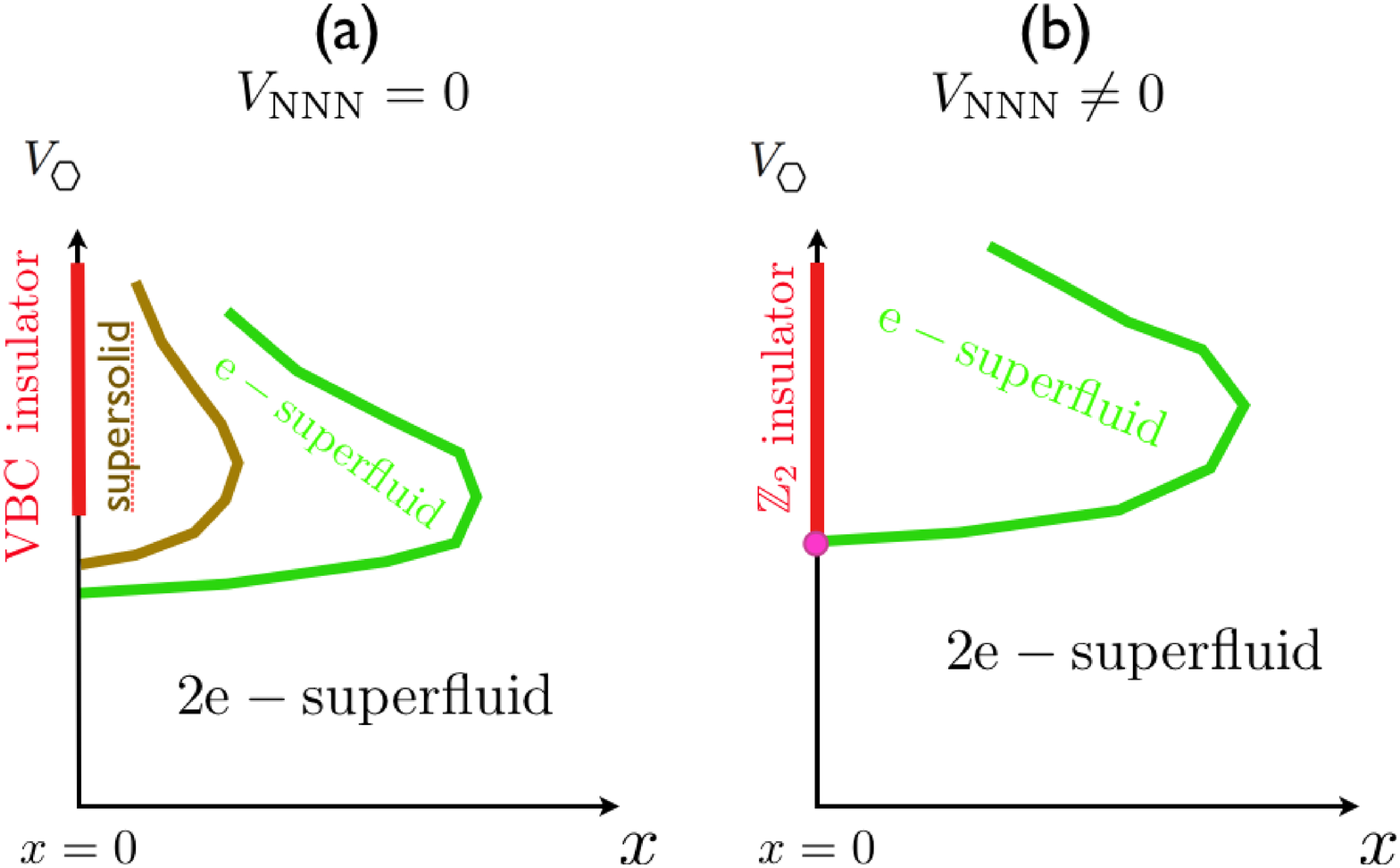}
\end{center}
\caption{ \label{fig:ph_diag_HCB} (color online) Schematic and
speculative phase diagram of interacting HCB on the kagome lattice for
$V_{dd}=0$ (a), with fine tuning of the third-nearest neighbor repulsion (b) -- see text.
In (a), the exact nature of the transitions between the
$2e$-superconductor and the VBC insulator at $x=0$ needs to be further
investigated.  In (b), the dot at $x=0$ might correspond to the XY$^*$
fractional critical point between the superfluid and the topological
$\mathbb{Z}_2$ insulator. The $e$-superfluid is the same phase as in the
doped QDM of Fig.~\ref{fig:phasediag_triangle}(a). }
\end{figure}

The mapping to the doped QDM on the triangular lattice is therefore
complete.  However, it is important to notice that $J=t^2/\Vhex >0$ (for
real $t$) so that only QDM (a) and (b) can be realized with HCB
depending on the sign of $t$. Introducing imaginary hoping $t=i\tau$ on
the kagome lattice equivalent to put U(1) fluxes through the triangles
lead to the QDM models (c) and (d) in the presence of a magnetic field.
In practice only the case of a real $t<0$ hopping on the original kagome
lattice can be handled with QMC.  Assuming the phase of the corresponding
doped QDM ((a) model) is a fractionalized charge-$e$ superfluid, we
therefore predict a non-conventional phase transition between the
(ordinary) $2e$-superfluid of the weakly interaction $d$-bosons and an
exotic $e$-superfluid at large $\Vhex/t$, as schematically shown in
Fig.~\ref{fig:ph_diag_HCB}. This is possible if the third-nearest neighbor repulsion is
carefully tuned -- $V_{\rm dd}\simeq J=t^2/\Vhex$. If $V_{\rm dd}=0$,
one gets a transition to a plaquette VBC phase at $x=0$, which might
involve intermediate phases. Indeed, close similarities are expected
with the melting of the (bosonic) plaquette VBC on the checkerboard
lattice, revealing an intermediate {\it commensurate}
supersolid.~\cite{Bose-Hubbard_checkerboard} Similarly to the effective
triangular QDM at $V=0$, doping of the VBC insulator should immediately
result into a supersolid phase which would melt into a charge-$e$
superfluid above some critical doping.  Lastly, at even larger doping
(corresponding to a dilute gas of dimers) a second phase transition to a
charge-$2e$ superfluid is expected.

\section{Conclusions}
\label{sec.conclusions}

In this paper we have established a rigorous and general equivalence
between QDM Hamiltonians with bosonic holes and a corresponding QDM
Hamiltonian with fermionic holes. Although this correspondence was
already noticed on the basis of numerical simulations in Ref.~\onlinecite{poilblanc_2008} and established 
analytically in Ref.~\onlinecite{stat1} for Hamiltonians with the
simplest flipping term, the correspondence has now been generalized to
more complicated cases. More importantly, we provide a general recipe to
very quickly -- and without any computation -- establish which are the two
equivalent Hamiltonians under this statistical transmutation.

We also note
that, when working with finite size systems, while the composite
particle representation is valid for any kind of boundary conditions, this is not 
the case for the method that use the Jordan-Wigner
transformation.  Indeed, the issue of boundary conditions 
in the two-dimensional version
of the Jordan-Wigner transformation has been very little discussed in
the literature. The point is that it does not seem possible to impose
periodic boundary conditions in a consistent way when using the
two-dimensional version of the Jordan-Wigner transformation, even if the
total numbers of particles is kept fixed.  This can be contrasted with
the one-dimensional version of the transformation where periodic boundary 
conditions can
be consistently imposed provided one keeps the number of particles
fixed.  In this sense the analytical results obtained \cite{stat1} with
the help of the Jordan-Wigner transformation are only valid to infinite
systems or finite system with open boundary conditions while the composite particle
representation used here can be consistently applied for any kind of 
boundary conditions. We have
then provided many examples of equivalent Hamiltonians for the more
generic cases of the square, triangular, hexagonal and kagome lattices.

In order to characterize the exotic superfluid phase with
condensation of fractionalized charge-$e$ holons, we
have introduced the gauge-invariant holon Green's function.
We have then considered four inequivalent cases of QDM Hamiltonians on
the triangular lattices and have numerically studied
various correlation functions including the above mentioned
holon Green's function, by Lanczos exact diagonalization of finite-size clusters.
We obtained rather strong and direct evidence for the existence of the exotic superfluid phase 
due to condensation of holons carrying charge $e$.
In fact, our numerical results suggest that all the four models we have studied
exhibit the charge-$e$ superfluid phase. More conventional charge-$2e$ and $4e$ 
superfluid phases are also present. 
While the existence of a charge-$e$ superfluid phase may be naturally understood
when the bare particle are bosons, it is
much more puzzling in the case where they are fermions, as it
corresponds to a superconductor without Cooper pairing.
Although one may expect that the condensation of charge $e$ holons
requires the holes to be bosons, our rigorous mapping
shows that fermionic statistics can be always assigned to holes
in the microscopic Hamiltonian.
This implies a dynamical statistical transmutation of holons in the
QDM where holes are represented as fermions. Those kind of dynamical statistical
transmutations can be monitored by studying the nodes of the wave function
and one spectacular example can be found in model (d) at intermediate
values of the doping where the system seems to switch from a Fermi liquid
to a (bosonic) $e$-superfluid phase. 

We have then provided a concrete microscopic
Bose-Hubbard Hamiltonian which in the strong interaction limit behaves as
a QDM on the triangular lattice, as the one analyzed numerically. It
allow us to have a better control on the doping (by simply varying the
number of bosons) and to better visualize the superfluid phases $e$-superfluid and $2e$-superfluid.
It is important to stress that a QDM may arise as effective low energy
models of quite different microscopic Hamiltonians. As such, the
physical consequences of various phases in the QDM can depend
on the mapping. For example, if the QDM arises from a
microscopic electronic Hubbard model, the holes are real electrons
vacancies (models (c) and (d)), and then they are charged. In this case the different
superfluid phases are superconducting phases. 
There is however another
way in which one could introduce doping. Imagine for example a system in
which there is no real electron vacancies but some magnetic field is
applied to the system. The effect of the magnetic field may have as
effect to break some of the singlet that are represented by the dimers
leaving two polarized spin $1/2$, which now play the role of the
holes~\cite{RBP} (models (a) and (b)). In this case 
the holes are neutral but carry spin, so that the
superfluid phase now corresponds to a superconductor of magnetic
current. Our results provide a validation of the 
previous claim~\cite{RBP} of an exotic superfluid of condensed deconfined and polarized spinons
(equivalent to our charge $e$-superfluid). In addition, we predict here the existence of another phase
of deconfined spinons, the Bose liquid, corresponding to an exotic spin 
liquid carrying {\it uncondensed} (polarized) spinons.
Interestingly, such exotic phases could indeed be realized in simple frustrated
magnetic systems, as for example the kagome anisotropic spin $1/2$ model
close to the magnetization plateau
at $1/3$ of the saturation value.\cite{lkagome1-3}.  
We hope that the results of this paper will establish
new motivations to investigate, with a new light, different
microscopic models which may give rise to the doped QDM as
an effective low energy model.

\section*{Acknowledgments}
We are grateful to D. C. Cabra with whom we initiated the work in this
subject. AR, DP and PP acknowledge support by the ``Agence Nationale de
la Recherche" under grant No.~ANR~2010~BLANC~0406-0.
MO is supported in part by Grant-in-Aid for Scientific Research
on Innovative Areas No. 20102008 from MEXT of Japan.
MO and DP were supported in part by the U.S. National Science
Foundation under Grant No. NSF PHY11-25915, while they were
at Kavli Institute for Theoretical Physics, UC Santa Barbara.
CAL is partially supported by CONICET (PIP 1691) and ANPCyT (PICT 1426).

\appendix

\section{Connection with the two-dimensional Jordan-Wigner transformation}

In this appendix we elaborate on an alternative proof of the statistical
transmutation in the QDM. In this approach we use a two dimensional
version of the Jordan-Wigner transformation. The fundamental ingredients
of this transmutations were presented for the square and triangular
lattices in a previous work \cite{stat1}.  It is important to stress
that this procedure is totally generic and can be implemented in any
two-dimensional lattice with open boundary conditions\cite{JW1,JW2}. As the main
technical steps for the square and triangular lattice were already
presented in \onlinecite{stat1}, here we only show the details for the
kagome lattice. As we have discussed above, for this lattice there is a
big freedom in choosing the ordering prescription.  This fact given rise
to an extra freedom in taking the sign of the flipping constants when we
change the statistics of holes.  We will see in this section that this
freedom is materialized whithin the Jordan-Wigner approach by using
gauge transformations on dimers and holes.
 Let us start with a quantum hard-core dimer model in the presence of
 holons on the Kagome lattice given by the following Hamiltonian:
\ba \label{eq:H_kagome} H=H_{J}+H_{V}+H_{t} \ea
The terms $H_{V}$ and $H_{J}$ corresponding to the diagonal and
 off-diagonal terms of the pure dimer model are taken up to resonance
 plaquettes of length 12 \cite{Ralko_Poilblanc_kagome,Zeng_1995}.
\begin{widetext}
\ba \label{eq:HJ_kagome} H_{J}\!\!&=&\!\!J_{6} \, \figu{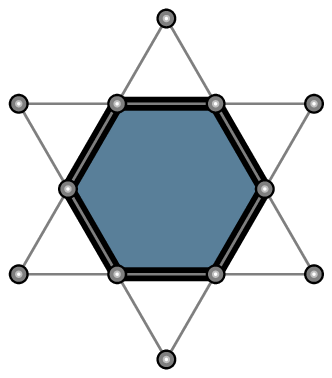}
+ J_{8}^{(a)} \figu{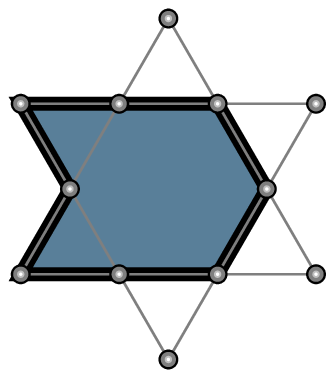} +
J_{8}^{(b)}\figu{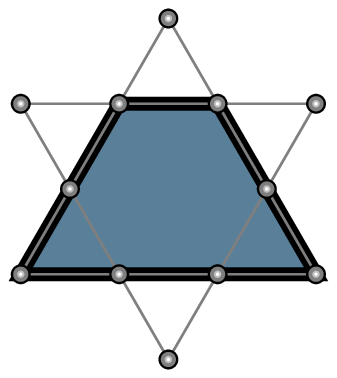} +J_{8}^{(c)}\figu{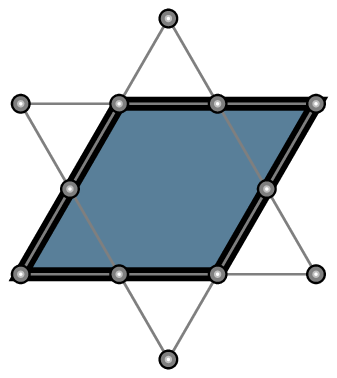}
+ J_{10}^{(a)}\figu{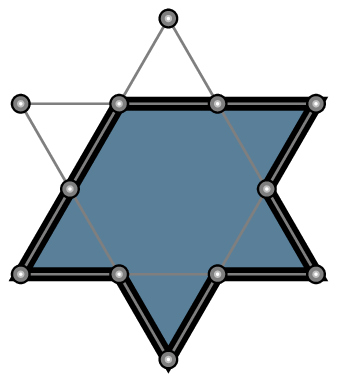} +
J_{10}^{(b)}\figu{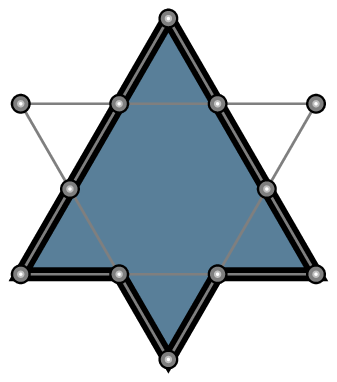}
+J_{10}^{(c)}\figu{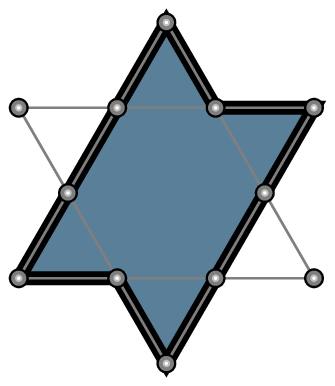} +J_{12} \,
\figu{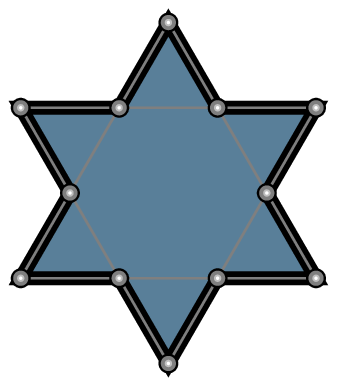}\\
\label{eq:HV_kagome} H_{V}\!\!&=&\!\!V_{6} \, \figu{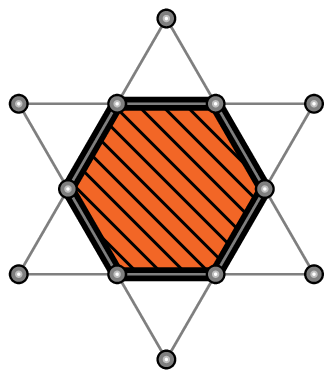} +
V_{8}^{(a)} \figu{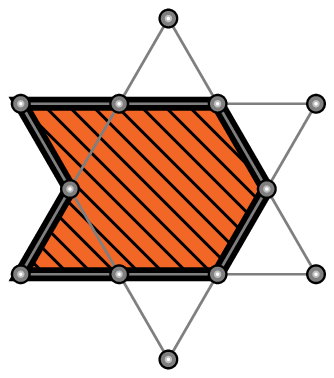} +
V_{8}^{(b)}\figu{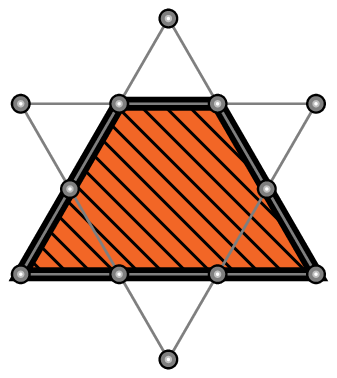}
+V_{8}^{(c)}\figu{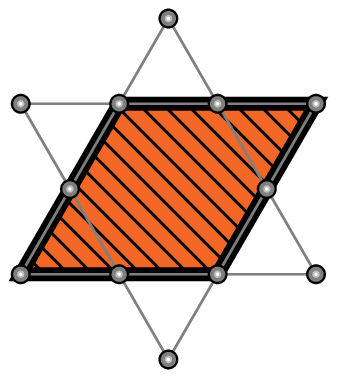} + V_{10}^{(a)}
\figu{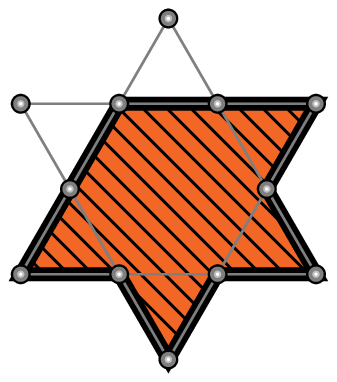} + V_{10}^{(b)} \figu{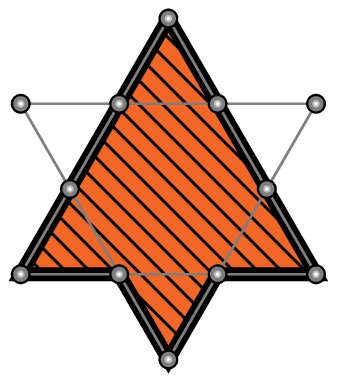}
+V_{10}^{(c)} \figu{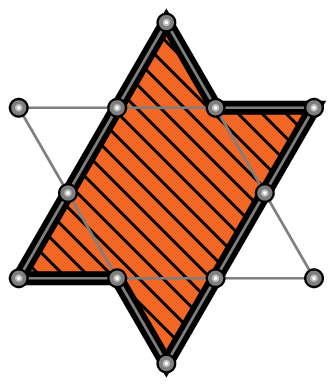} +V_{12} \,
\figu{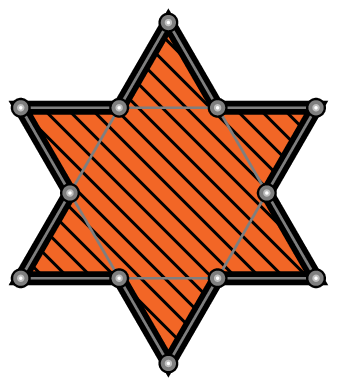}.  \ea
\end{widetext}
The sum runs over all the hexagons of the lattice and all the possible
 orientations of the plaquettes are implicit.  The kinetic and diagonal
 terms are given by \ba \figu{plaq_6.eps}&=&\sum_{\sumastar}\left\{
\Bigl| \,
\raisebox{-2.2ex}{\includegraphics[scale=0.25]{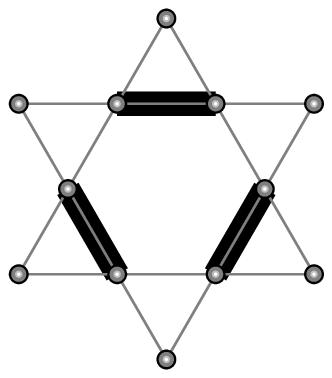}} \,
\Bigr\rangle
\Bigl\langle \,
\raisebox{-2.2ex}{\includegraphics[scale=0.25]{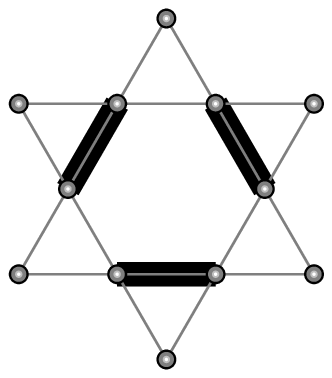}} \,
\Bigr|
%
 +\hbox{H.c.} \right\} \\\nn
\figu{V_plaq_6.eps}&=& \sum_{\sumastar} \left\{
\Bigl| \,
\raisebox{-2.2ex}{\includegraphics[scale=0.25]{J8_d1.eps}} \,
\Bigr\rangle
\Bigl\langle \,
\raisebox{-2.2ex}{\includegraphics[scale=0.25]{J8_d1.eps}} \,
\Bigr|
+ \Bigl| \,
\raisebox{-2.2ex}{\includegraphics[scale=0.25]{J8_d2.eps}} \,
\Bigr\rangle
\Bigl\langle \,
\raisebox{-2.2ex}{\includegraphics[scale=0.25]{J8_d2.eps}} \,
\Bigr|
\right\} \ea
and similar expressions for the other terms.  For convenience, the
 labels $\alpha$ in the amplitudes $J_{\alpha}$ and $V_{\alpha}$
 correspond to the length of the associated resonance loops and when it
 corresponds we add the label $(a)$, $(b)$ or $(c)$ corresponding to the
 3 non-equivalent plaquettes for the cases of length 8 and 10.  The
 Hamiltonian corresponding to the hoping of holons is given by
\ba H_{t}=H^{(t)}_{\vartriangle}+H^{(t)}_{\triangledown}, \ea
where
\ba \nonumber H^{(t)}_{\triangle}&=&-t\sum_{\vartriangle}\left\{
\figket{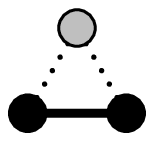}\figbra{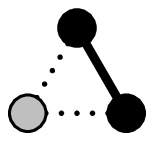} +
\figket{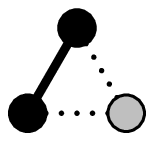}\figbra{h1.eps} +
\figket{h2.eps}\figket{h3.eps}\right.\\ &
+&\left. \hbox{H.c.}\right\}.  \ea
%
%
%
%

The hoping of holons can be written in a general way, independently of
the lattice, as a sum of three-site Hamiltonians
\ba H_{t}=\sum h^{(t)}_{(ijk)} \ea
with
\ba \label{eq:hoping_gral}
h^{(t)}_{(ijk)}=-t\;\hat{\mathcal{P}}\;b^{\dagger}_{i,j}b_{j,k}a^{\dagger}_{k}a_{i}\;\hat{\mathcal{P}}.
\ea
Where, we have been projected the Hamiltonian on the subspace where the
 constraint \ba \label{eq:constraint_kagome}
 a^{\dagger}_{i}a_{i}+\sum_{z}b^{\dagger}_{i,i+z}b_{i,i+z}=1, \ea
 is satisfied by mean the projectors $\hat{\mathcal{P}}$. Where the sum
 runs over NN of site $i$.
Starting from $h^{(t)}_{(i,j,k)}$, we transform the boson operators
$a_{i}$ using
\ba \label{eq:Jordan_Wigner} a_{i}=e^{-i\phi_{i}}f_{i} \ea
with
\ba \label{eq:phase_f} \phi_{i}=\sum_{j\neq i}f^{\dagger}_{j}f_{j}
\arg(\vec{\tau}_{j}-\vec{\tau}_{i}) \ea
togheter with the following transformation for the dimer operators
\ba \label{eq:b_tilde} \tilde{b}^{\dagger}_{i,j}&=&b^{\dagger}_{i,j}\,
e^{-i(\tilde{\phi}_{i}+\tilde{\phi}_{j})}\\ \tilde{b}_{i,j}&=&
e^{i(\tilde{\phi}_{i}+\tilde{\phi}_{j})} \, b_{i,j} \ea
 we obtain
\ba \label{eq:transformed_h_ijk_bis} h^{(t)}_{(i,j,k)}=-\tilde{t}\;
\hat{\mathcal{P}}\;\tilde{b}^{\dagger}_{i,j}\tilde{b}_{j,k}
f^{\dagger}_{k}f_{i}\;\hat{\mathcal{P}},
\ea
where the hoping amplitude is given by
\ba \label{eq:t_tilde}
\tilde{t}=t\,e^{i[\pi+\arg(\tau_{j}-\tau_{i})-\arg(\tau_{j}-\tau_{k})]}.
\ea

%
Equations (\ref{eq:transformed_h_ijk_bis}) and (\ref{eq:t_tilde}) are,
 in fact, independent of the lattice details. The information concerning
 the lattice geometry is contained in the arguments on the exponential
 of Eq. (\ref{eq:t_tilde}).  This equation can be written in a compact
 form as
\ba \tilde{t}=-t\,e^{i\psi_{t}}\, , \ea
where $\psi_{t}=\arg(\tau_{j}-\tau_{i})-\arg(\tau_{j}-\tau_{k})$ can be
represented graphically as \ba
\psi_{t}=\fifi[0.6]{-2.5}{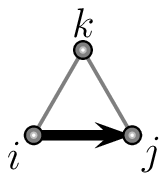}-\fifi[0.6]{-2.5}{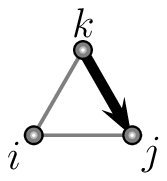}
\ea and $\fifi[0.6]{-2.5}{diag_t2.eps}$ represents the
$\arg(\vec{\tau}_{j}-\vec{\tau}_{i})$.

Now, we study the kinetic Hamiltonian corresponding to dimers.  Let us
start with the smallest resonance loop compatible with NN dimers on the
Kagome lattice, the plaquete of length 6. In this plaquette, the
resonance of the two possible dimerizations is given by
\ba H_{J_{6}}=J_{6}\sum_{\sumastar} \left\{
\Bigl| \,
\raisebox{-2.2ex}{\includegraphics[scale=0.25]{J8_d1.eps}} \,
\Bigr\rangle
\Bigl\langle \,
\raisebox{-2.2ex}{\includegraphics[scale=0.25]{J8_d2.eps}} \,
\Bigr|
 +\hbox{H.c.} \right\}.  \ea
In order to transform the dimers to the new representation using the
 flux generated by the statistical transformation of the holons we write
 the Hamiltonian in terms of dimer operators $b_{i,j}$ as
\ba H_{J_{6}}=\sum_{\sumastar} h^{(J_{6})}_{(i,j,k,l,m,n)} \ea
with
\ba h^{(J_{6})}_{(i,j,k,l,m,n)} = J_{6} \; b^{\dagger}_{i,j}
b^{\dagger}_{k,l} b^{\dagger}_{m,n} b_{j,k}b_{l,m}b_{n,i} + \hbox{H.c.}
\ea
Using the transformation (\ref{eq:b_tilde}) is straightforward to write
the Hamiltonian as
\ba h^{(J_{6})}_{(i,j,k,l,m,n)} = \tilde{J}_{6} \;
\tilde{b}^{\dagger}_{i,j} \tilde{b}^{\dagger}_{k,l}
\tilde{b}^{\dagger}_{m,n} \tilde{b}_{j,k}\tilde{b}_{l,m}\tilde{b}_{n,i}
+\hbox{H.c.}  \ea
where
\ba \tilde{J}_{6}=-J_{6}e^{i\psi_{6}} \ea and
\ba \nn \psi_{6}&=& \Bigl(
\arg(\vec{\tau}_{m}-\vec{\tau}_{n})+\arg(\vec{\tau}_{i}-\vec{\tau}_{j})+\arg(\vec{\tau}_{k}-\vec{\tau}_{l})
\Bigr)\\\nn
&-&\Bigl((\arg(\vec{\tau}_{n}-\vec{\tau}_{i})+\arg(\vec{\tau}_{j}-\vec{\tau}_{k})+\arg(\vec{\tau}_{l}-\vec{\tau}_{m})\Bigr).
\ea
It is convenient to use a graphical representation for the phase
$\psi_{6}$
\ba
\psi_{6}=\fifi[0.35]{-3.0}{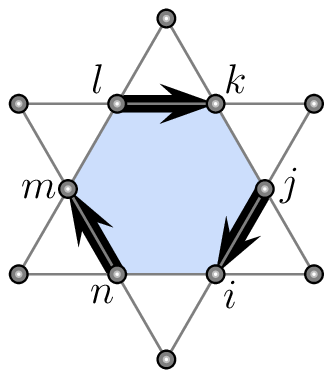}-\fifi[0.35]{-3.0}{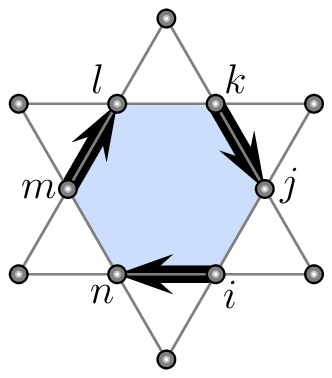}
\ea
where
\small \ba \nonumber \fifi[0.4]{-3.0}{diag_6_d1.eps}&=&\Bigl(
\arg(\vec{\tau}_{m}-\vec{\tau}_{n})+\arg(\vec{\tau}_{i}-\vec{\tau}_{j})+\arg(\vec{\tau}_{k}-\vec{\tau}_{l})
\Bigr)\\
\nonumber
\fifi[0.4]{-3.0}{diag_6_d2.eps}&=&\Bigl((\arg(\vec{\tau}_{n}-\vec{\tau}_{i})+\arg(\vec{\tau}_{j}-\vec{\tau}_{k})+\arg(\vec{\tau}_{l}-\vec{\tau}_{m})\Bigr)
\ea
\normalsize
The two graphs correspond to the initial and final dimerization on the
 plaquette.  In each graph we replace the dimers by arrows drawn in a
 clockwise direction and each graph represent the sum on the arguments
 of the arrows.

For the resonant plaquettes of length 8 we have 3 topologically distinct
 configurations.  Let us study now the term corresponding to the
 resonance plaquette $\fifi{-3.0}{plaq_8_2.eps}$.  After write it
 in terms of dimer operators and transform following (\ref{eq:b_tilde})
 we obtain
\ba \tilde{J}^{(b)}_{8}=-J^{(b)}_{8}e^{i\psi^{(b)}_{8}} \ea with
\ba
\psi^{(b)}_{8}=\fifi[0.35]{-3.0}{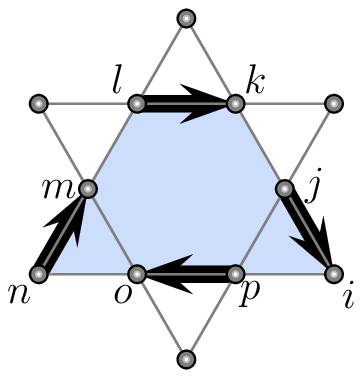}-\fifi[0.35]{-3.0}{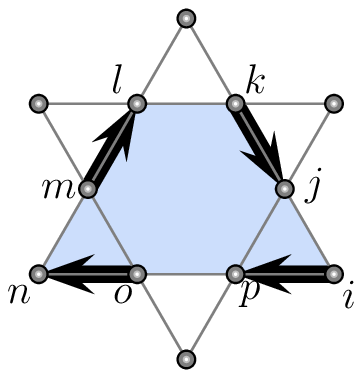}
\ea
%
%
where
\small \ba \nonumber \fifi[0.4]{-3.0}{diag_8_2_d1.eps}&=& \Bigl(
\arg(\vec{\tau}_{m}-\vec{\tau}_{n})+\arg(\vec{\tau}_{k}-\vec{\tau}_{l})+\arg(\vec{\tau}_{i}-\vec{\tau}_{j})\\\nn
&+&\arg(\vec{\tau}_{o}-\vec{\tau}_{p}) \Bigr)\\\nn
\fifi[0.4]{-3.0}{diag_8_2_d2.eps}&=&
\Bigl((\arg(\vec{\tau}_{n}-\vec{\tau}_{o})+\arg(\vec{\tau}_{l}-\vec{\tau}_{m})+\arg(\vec{\tau}_{j}-\vec{\tau}_{k})\\\nn
&+&\arg(\vec{\tau}_{p}-\vec{\tau}_{i})\Bigr) \ea \normalsize

It is easy to check that, under transformation (\ref{eq:b_tilde}), the
  amplitudes $J^{(\gamma)}_{\alpha}$ corresponding to the rest of the
  resonance plaquettes also transform as
\ba
\tilde{J}^{(\gamma)}_{\alpha}=-J^{(\gamma)}_{\alpha}e^{i\psi^{(\gamma)}_{\alpha}},
\ea {\it where $\psi^{(\gamma)}_{\alpha}$ is the phase obtained from the
difference between the two possible dimerizations in a given resonance
plaquette of the sum of the arguments corresponding to dimers oriented
clockwise (or anticlockwise).  }

This {\it graphical rule } can be used to study higher order therms in
the kinetic Hamiltonian.  This allows to determine the Hamiltonian after
the J-W transformation on the holons. In the Kagome lattice, up to
resonance plaquettes of length 12 we obtain that the Hamiltonian $H_{J}$
corresponding to kinetic energy of the dimers can be written as in
(\ref{eq:HJ_kagome}),
but dimers are now created by the operators $\tilde{b}^{\dagger}_{i,j}$
and the amplitudes $J_{\alpha}$ must be replaced by $\tilde{J}_{\alpha}$
where as the values of $V_{\alpha}$ remains unchanged.  In table
\ref{tab:kagome_diag} we show the values $\tilde{J}_{\alpha}/J_{\alpha}$
for the resonance plaquettes up to length 12.

\begin{table}
\begin{center}
\begin{tabular}{|c|c|c|c|c|}
\hline
\multicolumn{1}{|c|}{Length}&\multicolumn{1}{|c|}{6}&\multicolumn{3}{|c|}{8}\\
\hline
  & $\fifi[0.3]{-3.0}{plaq_6.eps}$ &
  $\fifi[0.3]{-3.0}{plaq_8_1.eps}$ &
  $\fifi[0.3]{-3.0}{plaq_8_2.eps}$ &
  $\fifi[0.3]{-3.0}{plaq_8_3.eps}$ \\\hline
$\psi$ & $\pi$ & $0$ & $\pi$ & $0$ \\\hline $\tilde{J}/J$ & 1 & -1 & 1 &
-1 \\\hline\hline
%
%
%
%
\multicolumn{1}{|c|}{Length}&\multicolumn{3}{|c|}{10}&\multicolumn{1}{|c|}{12}\\\hline
  & $\fifi[0.3]{-3.0}{plaq_10_1.eps}$ &
  $\fifi[0.3]{-3.0}{plaq_10_2.eps}$ &
  $\fifi[0.3]{-3.0}{plaq_10_3.eps}$ &
  $\fifi[0.3]{-3.0}{plaq_12.eps}$ \\\hline
$\psi$ & $\pi$ & $0$ & $\pi$ & $0$ \\\hline $\tilde{J}/J$ & 1 & -1 & 1 &
-1 \\\hline

\end{tabular}
\end{center}
\caption{Values of $\tilde{J}_{\alpha}/J_{\alpha}$ corresponding to the
lowest orders of the resonant plaquettes. } \label{tab:kagome_diag}
\end{table}

After the transmutation, the Hamiltonian corresponding to the hoping of
holes is given by
\ba
\tilde{H}_{t}=\tilde{H}^{(t)}_{\vartriangle}+\tilde{H}^{(t)}_{\triangledown}
\ea
where in $\tilde{H}^{(t)}_{\vartriangle}$ dimers are also created by
 operators $\tilde{b}^{\dagger}_{i,j}$ and the holes are fermions
 created by the operators $f^{\dagger}_{i}$:

\ba \nn
\tilde{H}^{(t)}_{\vartriangle}&=&-\tilde{t}\sum_{\vartriangle}\left\{
\figket{h1.eps}\figbra{h2.eps} +
\figket{h3.eps}\figbra{h1.eps} +
\figket{h2.eps}\figket{h3.eps}\right.\\
\label{eq:H_t_tilde_kagome} & +&\left. \hbox{H.c.}  \right\}, \ea
%
where $\tilde{t}=te^{-i\frac23 \pi}$. In order to obtain the original
value of the hoping constant for the holons ($\tilde{t}\rightarrow t$),
we can perform a gauge transformation on the dimers.  This is possible,
for example, by mean the following gauge transformation
\ba \nn
\Bigl|\,%
\raisebox{0.3ex}{\includegraphics[scale=0.3,angle=0]{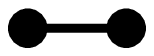}}\,%
\Bigr\rangle &\rightarrow & \Bigl|\,%
\raisebox{0.3ex}{\includegraphics[scale=0.3,angle=0]{dimer_solo.eps}}\,%
\Bigr\rangle
\\\nn
\Bigl|\,%
\raisebox{-0.9ex}{\includegraphics[scale=0.3,angle=60]{dimer_solo.eps}}\,%
\Bigr\rangle & \rightarrow & e^{-i\frac{2}{3}\pi} \Bigl|\,%
\raisebox{-0.9ex}{\includegraphics[scale=0.3,angle=60]{dimer_solo.eps}}\,%
\Bigr\rangle
\\\nn
\Bigl|\,%
\raisebox{-0.5ex}{\includegraphics[scale=0.3,angle=120]{dimer_solo.eps}}\,%
\Bigr\rangle & \rightarrow & e^{i\frac{2}{3}\pi} \Bigl|\,%
\raisebox{-0.5ex}{\includegraphics[scale=0.3,angle=120]{dimer_solo.eps}}\,%
\Bigr\rangle ,
\ea
independently if the dimers are on up or down triangles. We have for the
hole hoping term

\ba \nn \tilde{H}^{(t)}_{\vartriangle}&=&-t\sum_{\vartriangle}\left\{
\figket{h1.eps}\figbra{h2.eps} +
\figket{h3.eps}\figbra{h1.eps} +
\figket{h2.eps}\figket{h3.eps}\right.\\
\label{eq:H_t_tilde_kagome_2} & +&\left. \hbox{H.c.}  \right\} \ea
while for the resonance terms the gauge transformation does not change
 the couplings $\tilde{J}$ and $V$.  Then we obtain
\ba \nn \tilde{H}_{J}&=&J_{6} \, \figu{plaq_6.eps} -
J_{8}^{(a)}\figu{plaq_8_1.eps} +
J_{8}^{(b)}\figu{plaq_8_2.eps} -J_{8}^{(c)}\figu{plaq_8_3.eps}
\\\nn &+& J_{10}^{(a)}\figu{plaq_10_1.eps} -
J_{10}^{(b)}\figu{plaq_10_2.eps}
+J_{10}^{(c)}\figu{plaq_10_3.eps} -J_{12} \,
\figu{plaq_12.eps}
\ea
Finally we can use another gauge transformation to get a more simple
Hamiltonian. We change the dimers on up triangles as
\ba \nn
\Bigl|\,%
\raisebox{-0.7ex}{\includegraphics[scale=0.4,angle=0]{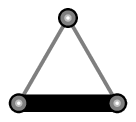}}\,%
\Bigr\rangle &\rightarrow & e^{i\theta_{1}} \Bigl|\,%
\raisebox{-0.7ex}{\includegraphics[scale=0.4,angle=0]{dim1up.eps}}\,%
\Bigr\rangle
\\\nn
\Bigl|\,%
\raisebox{-0.7ex}{\includegraphics[scale=0.4,angle=0]{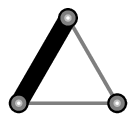}}\,%
\Bigr\rangle & \rightarrow & e^{i\theta_{2}} \Bigl|\,%
\raisebox{-0.7ex}{\includegraphics[scale=0.4,angle=0]{dim2up.eps}}\,%
\Bigr\rangle
\\\nn
\Bigl|\,%
\raisebox{-0.7ex}{\includegraphics[scale=0.4,angle=0]{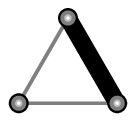}}\,%
\Bigr\rangle & \rightarrow & e^{i\theta_{3}} \Bigl|\,%
\raisebox{-0.7ex}{\includegraphics[scale=0.4,angle=0]{dim3up.eps}}\,%
\Bigr\rangle
\ea
while we change the corresponding to down triangles as
\ba \nn
\Bigl|\,%
\raisebox{1.9ex}{\includegraphics[scale=0.4,angle=180]{dim1up.eps}}\,%
\Bigr\rangle &\rightarrow & e^{i\varphi_{1}} \Bigl|\,%
\raisebox{1.9ex}{\includegraphics[scale=0.4,angle=180]{dim1up.eps}}\,%
\Bigr\rangle
\\\nn
\Bigl|\,%
\raisebox{1.9ex}{\includegraphics[scale=0.4,angle=180]{dim2up.eps}}\,%
\Bigr\rangle & \rightarrow & e^{i\varphi_{2}} \Bigl|\,%
\raisebox{1.9ex}{\includegraphics[scale=0.4,angle=180]{dim2up.eps}}\,%
\Bigr\rangle
\\\nn
\Bigl|\,%
\raisebox{1.9ex}{\includegraphics[scale=0.4,angle=180]{dim3up.eps}}\,%
\Bigr\rangle & \rightarrow & e^{i\varphi_{3}} \Bigl|\,%
\raisebox{1.9ex}{\includegraphics[scale=0.4,angle=180]{dim3up.eps}}\,%
\Bigr\rangle
\ea
Taking the values $\theta_{1}=\theta_{2}=\varphi_{1}=\varphi_{2}=0$,
$\theta_{3}=\pi / 2$ and $\varphi_{3}=-\pi / 2$. The amplitudes
$J_{\alpha}$ change as
\ba \nn \tilde{H}_{J}&=&-J_{6} \, \figu{plaq_6.eps} -
J_{8}^{(a)}\figu{plaq_8_1.eps} -
J_{8}^{(b)}\figu{plaq_8_2.eps} -J_{8}^{(c)}\figu{plaq_8_3.eps}
\\\nn &-& J_{10}^{(a)}\figu{plaq_10_1.eps} -
J_{10}^{(b)}\figu{plaq_10_2.eps}
-J_{10}^{(c)}\figu{plaq_10_3.eps} -J_{12} \,
\figu{plaq_12.eps}
\ea
Then finally we obtain a Hamiltonian with a global change of sign in the
amplitudes of the kinetic term.  After the transformation, the terms
corresponding to the hoping of holons becomes \ba \nn
\tilde{H}^{(t)}_{\vartriangle}&=&-t\sum_{\vartriangle}\left\{
e^{-i\pi/2}\figket{h1.eps}\figbra{h2.eps}  \right. \\ \nonumber &+& \left. 
\figket{h3.eps}\figbra{h1.eps} +e^{i\pi/2}
\figket{h2.eps}\figket{h3.eps}\right\} \\ &+& \hbox{H.c.}
\ea
This hoping Hamiltonian can be transformed by mean a simple gauge
 transformation on the holes, $f_{j}\rightarrow
 e^{i\vec{Q}\cdot\vec{\tau}_{j}}f_{j}$ with
 $\vec{Q}=(\frac{\pi}{2},\frac{\pi}{2\sqrt{3}})$, to recover the
 original form of the Hamiltonian.
%
 With this transformation we have that
\ba \nn f^{\dagger}_{j+p_{1}}f_{j}&\rightarrow & e^{i\frac{\pi}{2}}
f^{\dagger}_{j+p_{1}}f_{j}\\\nn f^{\dagger}_{j+p_{2}}f_{j}&\rightarrow &
e^{i\frac{\pi}{2}} f^{\dagger}_{j+p_{2}}f_{j}, \ea
and we recover the original form (\ref{eq:H_t_tilde_kagome_2}) for the
 Hamiltonian corresponding to the holes.  Then, finally we have that
 changing the statistics of holons together with the sign of all the
 kinetic amplitudes we obtain a completely equivalent Hamiltonian.

By using the J-W transformation we can recover the equivalences
 presented in previous section, but the procedure is more laborious. We
 have pointed that starting from different prescriptions for the bonds
 we can prove different equivalences. A change i the lattice
 prescription used in the composite operator representation is
 equivalent to a gauge transformation on on the dimers.

One last very important point concerns the issue of boundary conditions
within the J-W approach.  The composite operator approach which we have
extensively used in the paper is valid independently of the boundary
conditions used for the system. In contrast, the J-W approach is valid
only for infinite systems or finite systems with open boundary
conditions.  Let us come back to equation (\ref{eq:Jordan_Wigner})
assuming that the bosonic $a$ operators are well defined in a system in
periodic boundary conditions.  This means that for example the operators
$a_{i}$ and $a{\tilde{i}}$ are forced to be the same if sites $i$ and
$\tilde{i}$ correspond to the same point in the system because of the
periodic boundary conditions. However, because of the very nature of the
non-local J-W transformation, the relation between the operators $f_{i}$
and $f{\tilde{i}}$ must contain a twist. In the one-dimensional version
of the J-W transformation, it is easy to see that this twist is a sign
which depends only on the total number of particles in the system. Then,
in one dimension, if one restricts to the subspace of a fixed number of
particles it is possible to implement consistent way periodic boundary
conditions. Here in two dimensions, the twist one should force for the
fermionic operators depends not only on the number of particles but also
on their relative positions with respect to the points $i$ and
$\tilde{i}$. That means that, even restricting to a fixed number of
particles, starting from periodic boundary conditions for the bosons
operators, it is not possible to implement boundary conditions for the
fermionic operators which are consistent with all the possible particle
configurations.

%
%
%
%

\end{document}